\documentclass[aps,prd,nofootinbib,preprint,superscriptaddress]{revtex4}  
\usepackage{epsfig}
\usepackage{amsmath}

\begin{document}

\newcommand{\beq}{\begin{equation}}
\newcommand{\eeq}{\end{equation}}

\title{Predictions for $b\to ss\bar d$ and $b\to dd\bar s$ decays in the SM and
with new physics}
\preprint{CERN-PH-TH/2009-137}
\author{Dan Pirjol} 
\affiliation{National Institute for Physics and Nuclear Engineering, 
Department of Particle Physics, 077125 Bucharest, Romania}

\author{Jure Zupan}
\affiliation{Theory Division, Department of Physics, CERN, CH-1211 Geneva 23, Switzerland}
\altaffiliation{On leave of absence from Faculty of mathematics and physics, University of
  Ljubljana, Jadranska 19, 1000 Ljubljana, Slovenia, and Josef Stefan Institute, Jamova 39, 
  1000 Ljubljana, Slovenia}

\date{\today}

\begin{abstract}
The $b\to ss\bar d$ and $b\to dd\bar s$ decays are highly suppressed in the SM,
and are thus good probes of new physics (NP) effects. We discuss in detail the
structure of the relevant SM effective Hamiltonian pointing out the presence of nonlocal 
contributions which can be about $\lambda^{-4} (m_c^2/m_t^2) \sim $ 30\% of the local 
operators ($\lambda = 0.21$ is the Cabibbo angle). 
The matrix elements of the local operators are computed
with little hadronic uncertainty by relating them through
flavor SU(3) to the observed $\Delta S = 0$ decays. We identify a general NP mechanism 
which can lead to the branching fractions of the  $b\to ss\bar d$ modes at or just 
below the present experimental bounds, 
while satisfying the bounds from $K-\bar K$ and $B_{(s)}-\bar B_{(s)}$ mixing.
It involves the exchange of a NP field carrying a conserved charge,
broken only by its flavor couplings. The size of branching fractions
within MFV, NMFV and general flavor violating NP are also predicted. We show
that  in the future energy scales higher than $10^3$ TeV could be probed without
hadronic uncertainties even for $b\to s$ and $b\to d$ transitions, if enough
statistics becomes available.
\end{abstract}


\maketitle

\section{Introduction}
The decays $b\to ss\bar d $ and $b\to dd \bar s$ are highly suppressed in the SM: 
they are both loop and CKM suppressed (by six powers of small CKM elements $V_{ts}$ 
and/or $V_{td}$). 
As such they can be used for searches of New Physics (NP) signals 
\cite{Huitu:1998vn,Grossman:1999av,Fajfer:2000ny,Fajfer:2006av,Fajfer:2001ht,Cai:2004mi,Fajfer:2004fx,Wu:2003kp}.
The types of NP that would generate $b\to ss\bar d $ and $b\to dd \bar s$ transitions will commonly also give 
contributions to $K-\bar K$, $B-\bar B$ and $B_s-\bar B_s$ mixing. Since no clear deviations from the SM predictions 
are seen in the meson mixing, is it possible to have deviations in $b\to ss\bar d $ and $b\to dd \bar s$ transition 
observable at Belle~II and at LHCb? A related question is: with improved statistics, can the experiments using 
$b\to ss\bar d $ and $b\to dd \bar s$ decays push the bounds on flavor violation scale beyond what can be achieved 
from the mixing observables?

We address the second question first. For simplicity let us assume that NP contributions can be matched onto the
SM operator basis, so that $H^{\Delta S}=C_1^{sd}(\bar d_L \gamma^\mu s_L) (\bar d_L \gamma_\mu s_L)$, 
$H^{\Delta B}= C_1^{bs}(\bar s_L \gamma^\mu b_L) (\bar s_L\gamma_\mu b_L)$ and 
$H^{b\to ss\bar d}=C_1^{b\to ss\bar d} (\bar s_L\gamma^\mu b_L) (\bar s_L\gamma_\mu d_L)$. Using $|C_1^i|=1/(\Lambda^i)^2$ 
one finds \cite{Bona:2007vi}
\beq
\begin{split}
K-\bar K~{\rm mixing}:&\qquad \Lambda^{sd}> 1.0 \cdot 10^3~{\rm TeV},\\
B_d-\bar B_d~{\rm mixing}: &\qquad \Lambda^{bd}> 210~{\rm TeV},\\
B_s-\bar B_s~{\rm mixing}: &\qquad \Lambda^{bs}> 30~{\rm TeV},
\end{split}\label{bounds}
\eeq
with $\text{Im}(C_1^{sd})$ additionally constrained from $\varepsilon_K$.  
The above bounds should be compared with the following prediction for the $b\to ss\bar d$ 
transition in the presence of NP with scale  $\Lambda^{b\to ss\bar d}$ 
(see section~\ref{NPsection} for derivation)
\beq
{\cal B}(\bar B^0\to \bar K^{0*} \bar K^{0*})= 0.3 \times 10^{-6} \Big(\frac{10~{\rm TeV}}{\Lambda^{b\to ss\bar d}}\Big)^4,
\eeq
while the SM prediction for this branching ratio is of $\mathcal{O}(10^{-15})$. 
Let us take as an estimate $\Lambda^{b\to ss\bar d}\sim \sqrt{\Lambda^{bs} \Lambda^{sd}}$, 
a relation that holds in a wide set of NP models including the
Minimal Flavor Violation (MFV) and Next-to-Minimal Flavor Violation (NMFV) frameworks. 
With enough 
statistics the bound on $\Lambda^{bs}$ can then be pushed up to $10^3$ TeV and higher 
without running into SM background. The $b\to ss\bar d$ decay modes could thus be used to constrain the NP flavor 
structure for $b\to s$ transitions  as precisely as it is possible for $s\to d$ transitions from kaon physics. 
However, the statistics needed is very large. For instance, even to probe this type of flavor violating NP beyond 
the mixing bounds, the LHCb and Belle II luminosities
will not be enough. In this
scenario the $K-\bar K$ and $B_d-\bar B_d$ mixing bounds translate to 
${\cal B}(b \to dd\bar s)\lesssim 10^{-13}$ and the bounds from $K-\bar K$ and $B_s-\bar B_s$ mixing translate to 
${\cal B}(b \to ss\bar d)\lesssim 10^{-11}$.

Does this mean that any NP discoveries using $b \to dd\bar s$ and $b \to ss\bar d$ transitions are excluded at Belle~II and LHCb? Certainly not. It is possible to have significant effects in $b \to dd\bar s$ and $b \to ss\bar d$ while obeying the 
bounds from the meson mixing, if (i) the exchanged particle (or a set of particles) $X$  carries 
an approximately conserved global charge and,  if (ii) additionaly there  is some 
hierarchy in the couplings (or alternatively some cancellations in $K-\bar K$ mixing). 
Consider the NP Lagrangian of a generic form
\begin{eqnarray}\label{HeffX}
{\cal L_{\rm flavor}}=g_{b\to s}(\bar s \Gamma b)X + g_{s\to b} (\bar b \Gamma s)X
+ g_{d\to s}(\bar s \Gamma d)X + g_{s \to d} (\bar d \Gamma s)X + \mbox{h.c.},
\end{eqnarray}
and assume that $X$ carries a conserved quantum number broken only by the above terms. We also assume for simplicity that the field $X$ couples to a fixed Dirac structure $\Gamma$.
Integrating out the field $X$  produces flavor-changing operators
\beq\label{NPX}
\begin{split}
{\cal L}_{\rm eff} = \frac{1}{M_X^2} \big[&g_{ d\to s}g_{s\to d}^* (\bar s \Gamma d)(\bar s \bar \Gamma  d) +   g_{b\to s}g_{s\to b}^* (\bar s \Gamma b)(\bar s \bar \Gamma b)  \\
&+  g_{ b\to s}g_{s\to d}^*
(\bar s \Gamma b)(\bar s \bar \Gamma d) + 
 g_{d\to s}g_{s\to b}^* (\bar s \bar \Gamma b)(\bar s \Gamma d) \big],
\end{split}
\eeq
with the terms in the first line contributing to $K-\bar K$ mixing and $B_s - \bar B_s$ mixing, and in the second line to
$b \to ss\bar d$ decays (we also introduced $\bar \Gamma=\gamma^0 \Gamma^\dagger \gamma^0$). It is now possible to set contributions to meson mixing to zero, while keeping $b\to ss\bar d$
unbounded. This   happens for instance, if 
\beq\label{g:hierarchy}
g_{b\to s}\ll g_{s\to b}, \quad g_{s\to d}\ll g_{d\to s},\qquad\text{or}\qquad g_{b\to s}\gg g_{s\to b}, \qquad g_{s\to d}\gg g_{d\to s}.
\eeq
In this way all the present experimental bounds can be satisfied, while branching ratios for $b\to ss\bar d$ and 
$b\to dd\bar s$ induced decays are $ {\mathcal O}(10^{-6}) $ (see section \ref{NPsection} for details).

 The important ingredient in the above argument was that $X$ carried a conserved quantum number, so that there were no terms in ${\cal L}_{\rm eff}$  of the form
 \beq
\begin{split}\label{generic}
 g_{b\to s}^2(\bar s\Gamma b)(\bar s\Gamma b)+
g_{d\to s}^2(\bar s \Gamma d)(\bar s \Gamma d)+g_{s\to b}^{*2}(\bar s\bar \Gamma b)(\bar s\bar \Gamma b)+
g_{s\to d}^{*2}(\bar s \bar \Gamma d)(\bar s \bar \Gamma d)\dots,
\end{split}
\eeq
These would be generated for $X=X^\dagger$, which is impossible, if $X$ carries a conserved
charge. If terms \eqref{generic} are present, then
 $B_s-\bar B_s$ mixing forces both $g_{b\to s}$ and $g_{s \to b}$ to be small, 
and the hierarchy in \eqref{g:hierarchy} is not possible (similarly $K-\bar K$ mixing bounds $g_{d\to s}$ and $g_{s\to d}$ to both be small). 
An explicit example of a NP scenario where only terms of the form  \eqref{NPX} are generated is 
$R-$parity violating MSSM~\cite{Fajfer:2006av}. 
The $R$-parity violating term in the superpotential, $W=\lambda_{ijk}'L_i Q_j\bar d_k$, leads to 
$\tilde \nu_i \bar q_{Lj} d_{kR}$ flavor violating coupling. Sneutrino exchange generates operators of the form 
\eqref{NPX}, while operators of the form \eqref{generic} are not
generated, since the sneutrino carries lepton charge broken only by $R$-parity violating terms. 

A hierarchy of couplings in \eqref{g:hierarchy} is also present in (N)MFV models, if left-right terms 
give dominant contributions \cite{Kagan:2009bn}.
Both terms in \eqref{NPX} and \eqref{generic} are generated, on the other hand, for FCNCs induced by $Z'$ exchange, 
since $Z'$ does not carry any conserved charge.

In this paper we will not confine ourselves to a particular model but keep the analysis completely general 
using effective field theory. 
We will improve on the existing SM predictions, and also give predictions for general NP contributions. 
The most general local NP hamiltonian for $b\to ss\bar d$ transition is \cite{Grossman:1999av}
\begin{eqnarray}\label{HeffNP}
H^{\rm NP}=\frac{1}{\Lambda_{\rm NP}^2}\Big(\sum_{j=1}^5 c_j Q_j + \sum_{j=1}^5 \tilde c_j\tilde Q_j\Big),
\end{eqnarray}
where $c_j$ are dimensionless Wilson coefficients, $\Lambda_{\rm NP}$ the NP scale, and the operators are
\beq
\begin{split}\label{Qops}
Q_1=&(\bar s_L \gamma_\mu b_L)(\bar s_L \gamma^\mu d_L), \\
Q_2=&(\bar s_R b_L) (\bar s_R d_L), \quad Q_3=(\bar s_R^\alpha b_L^\beta) (\bar s_R^\beta d_L^\alpha),\\
Q_4=&(\bar s_R b_L) (\bar s_L d_R), \quad Q_5=(\bar s_R^\alpha b_L^\beta) (\bar s_L^\beta d_R^\alpha).
\end{split}
\eeq
The $\tilde Q_j$ operators are obtained from $Q_j$ by $L\leftrightarrow R$ exchange. 
In SM only $Q_1$ is present. The $b \to dd\bar s$ effective Hamiltonian is
obtained by exchanging $s\leftrightarrow d$ in the above equations, while the $K-\bar K$ and $B_s-\bar B_s$ mixing Hamiltonians follow from  $b\to d$ and $d\to s$ replacements.

The predictions following from the NP hamiltonian (\ref{HeffNP}) require calculating the QCD matrix
elements of the four-quark operators. In this paper we show that the $Q_1$ matrix elements 
can be related by SU(3) flavor symmetry to linear combinations of observable 
$\Delta S = 0$ decay amplitudes. This gives clean predictions for the
branching fractions of the exclusive $b\to ss\bar d$ and $b\to dd\bar s$ modes in the SM and the NP models 
where $Q_1$ dominate. This happens in 
a large class of NP models, including the two-Higgs doublet model with small
$\tan\beta$, and the MSSM with
conserved R parity \cite{Fajfer:2006av}.  The effects of the operators with non-standard
chirality can be estimated using factorization.

The outline of the paper is as follows. In Section \ref{EffectiveHamiltonian} we review the structure of the
effective Hamiltonian mediating the $b\to ss\bar d, dd\bar s$ decays in the Standard Model.
We point out that in addition to the local operators, the effective Hamiltonian contains
also nonlocal operators which have not been included in the previous literature. 
In Section \ref{Sec:SU3} we derive the flavor SU(3) relations for the matrix elements of the $Q_1$ operator. 
The resulting numerical predictions for 
$b\to ss\bar d, dd\bar s$ decays in the SM are given in Section \ref{SMpredictions}. 
NP predictions in the case of $Q_1$ operator dominance are discussed in Section \ref{NPsection}, 
while in Section \ref{non-SM-chiral} the modifications needed for a general chiral structure are given. 
Three appendices contain further technical details. 

\section{SM Effective Hamiltonian for $b\to ss\bar d$ and $b\to dd\bar s$ decays}\label{EffectiveHamiltonian}
\begin{figure}[t] 
   \centering
\includegraphics[width=5in]{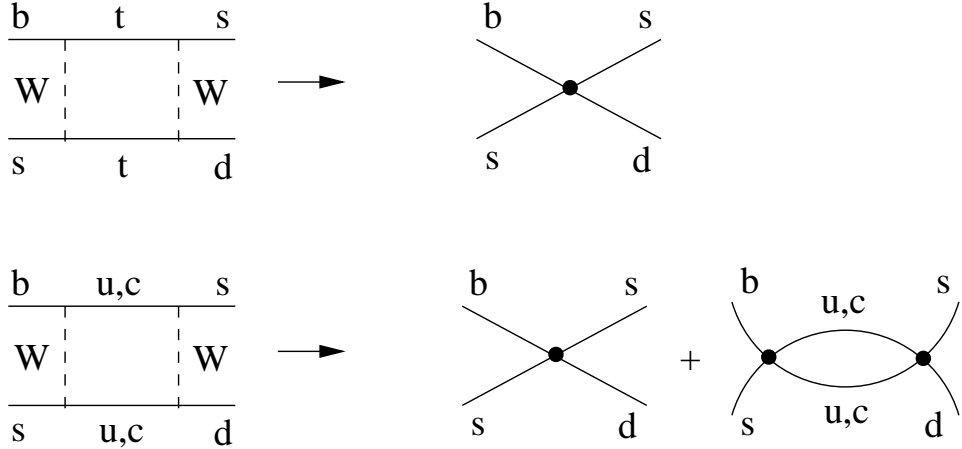} 
   \caption{Matching the box diagrams contributing to $b\to ss\bar d$ decays onto an
effective theory with $m_W \geq \mu \geq m_b$. The top quark box diagram (above) is matched onto
a local four-quark operator, while the box diagrams with $u,c$ internal quarks (below) are matched
onto local and nonlocal operators. The mixed top-charm and top-up box diagrams are power suppressed
by $m_{u,c}^2/m_W^2$ and do not contribute at leading order, see Appendix A.}
   \label{fig:matching}
\end{figure}

In the SM the $b\to ss\bar d, dd\bar s$ decays are mediated by the box diagram with internal $u,c,t$ quarks, Fig.~\ref{fig:matching}. For notational simplicity let us focus on the case of $b\to ss\bar d$, 
while the 
results for  $b\to dd\bar s$ can be obtained through a replacement $s\leftrightarrow d$.
The effective weak Hamiltonian for $b\to ss\bar d$ is obtained in analogy to the one for
$K^0-\bar K^0$ mixing \cite{Buchalla:1995vs,Witten:1976kx,Gilman:1982ap,Herrlich:1996vf}, but with
several important differences. First, the CKM structure is more involved. Second, the presence
of the massive $b$ quark in the initial state introduces a correction, which is however suppressed
by $m_b^2/m_W^2$, and is thus numerically negligible.
Finally, in applications to $K^0-\bar K^0$ mixing the charm quark can
be integrated out of the theory, while this cannot be done for exclusive $B$ decays, where 
there is no clear separation between the charm mass $m_c$ and the energy scales relevant in 
nonleptonic exclusive B decays into two pseudoscalars. 

At scales $m_b \leq \mu \leq m_W$, the effective weak Hamiltonian mediating $b\to ss\bar d$ 
decays contains both 
local $\Delta S=2$ terms as well as nonlocal terms arising from T-products of $\Delta S=1$ effective weak Hamiltonians
\begin{eqnarray}\label{HeffSM}
 {\cal H}_{ss\bar d} &={\cal H}^{\Delta S=2} +
\int d^dx \,T\big\{{\cal H}^{\Delta S=1}_d(x), {\cal H}^{\Delta S=1}_b(0)\big\}\,.
\end{eqnarray}
The local  part is
 \beq\label{DeltaS=2}
 {\cal H}^{\Delta S=2} =\frac{G_F^2 m_W^2}{16 \pi^2}\big(\lambda_t^d\lambda_t^b C_{tt}
 + \lambda_c^d\lambda_t^b C_{ct}+\lambda_t^d\lambda_c^b C_{tc}\big)\big[(\bar s d)_{V-A} (\bar s b)_{V-A}\big],
 \eeq
where the CKM structures are defined as $\lambda_q^{q'}=V_{qq'}V_{qs}^*$. 
The Wilson coefficient coming from the top box loop is $C_{tt}\sim O(x_t)$ and from 
the top-charm box loop $C_{ct}=C_{tc}\sim O(x_c)$, 
where $x_i=m_i^2/m_W^2$. The scaling of the three 
contributions in the local Hamiltonian \eqref{DeltaS=2} in terms of Cabibbo angle $\lambda=0.22$ 
and quark masses is then: 
$\sim\lambda^7 x_t$, $\sim\lambda^3 x_c$ and $\sim\lambda^7 x_c$ (for $b\to dd\bar s$ all three 
terms are suppressed by another factor of $\lambda$). 
The third term in  \eqref{DeltaS=2} can thus easily be neglected. Note also that there is no 
$\lambda_c^d\lambda_c^b$ term. The resulting absence of large $\log x_c$ from the charm box contribution is sometimes called the
{\it super-hard} GIM mechanism \cite{Herrlich:1996vf}, and follows from the chiral
structure of the weak interaction in the SM, as explained
in Appendix A.  The precise values of the Wilson coefficients in \eqref{DeltaS=2}
can be read off from the expressions for $K^0-\bar K^0$ mixing \cite{Buchalla:1995vs}, 
where the RG running is performed
only down to scale $\mu\sim m_b$. For $\bar m_t(\bar m_t)=160.9$ GeV, $\bar m_c(\bar m_c)=1.27$ GeV, $\alpha_s(m_Z)=0.118$ 
they are at $\mu=m_b=4.2$ GeV: $C_{tt}(m_b)=1.92$, $C_{tc}(m_b)=3.75 x_c=9.35 \cdot 10^{-4}$ at 
leading order (LO) (see appendix \ref{Wilson_coeffs} for the derivation).

The nonlocal contributions in $b \to ss\bar d$ transition, Eq. (\ref{HeffSM}), come from insertions of $\Delta S=1$ effective weak Hamiltonians ${\cal H}^{\Delta S=1}_b$ and ${\cal H}^{\Delta S=1}_d$ (see also Fig.~\ref{fig:matching}). The $\Delta S=1$ effective weak Hamiltonian ${\cal H}^{\Delta S=1}_b$ is  the same weak 
Hamiltonian relevant for hadronic $B$ decays
\beq\label{H_b}
 {\cal H}^{\Delta S=1}_b=\frac{G_F}{\sqrt2}\Big(\sum_{q,q'=u,c}V_{qs}^*V_{q'b}\sum_{i=1,2}C_i Q_{i,b}^{qq'} -V_{ts}^*V_{tb}\sum_{j=3}^6 C_jQ_j^b\Big),
 \eeq
with the tree operators $Q_{1,b}^{qq'}=(\bar qb)_{V-A} (\bar s q')_{V-A},\, Q_{2,b}^{qq'}=(\bar q_\beta b_\alpha)_{V-A} (\bar s_\alpha q'_\beta)_{V-A}$,
and penguin operators $Q_{3,5}^b=(\bar s b)_{V-A} (\bar q q)_{V\mp A },\,Q_{4,6}^b=(\bar s_\alpha b_\beta)_{V-A} (\bar q_\beta q_\alpha)_{V\mp A}$,
where the color indices $\alpha, \beta$  are displayed only when the sum is over the fields in 
different brackets. In the definition of the penguin operators $Q_{3-6}^b$ a sum over $q=\{u,d,s,c,b\}$ is implied. The weak Hamiltonian ${\cal H}^{\Delta S=1}_d$ follows 
from \eqref{H_b} by making the replacement $b\to d$.

Using CKM unitarity we can rewrite the CKM factors as $V_{us}^*V_{ub}=-V_{ts}^*V_{tb}-V_{cs}^*V_{cb}$. 
The insertions of tree operators with $u$ and $c$ quarks will generate  contributions with CKM 
structure $\lambda_c^d\lambda_c^b$, that are not present in the local $\Delta S=2$ Hamiltonian 
\eqref{DeltaS=2},
\beq\label{nonlocal-cc}
\begin{split}
{\cal H}_{cc}=\frac{G_F^2}{2}\lambda_c^d\lambda_c^b \int d^dx \sum_{i,j=1,2}C_iC_j& T\big\{Q_{i,d}^{cc}(x)Q_{j,b}^{cc}(0)+Q_{i,d}^{uu}(x)Q_{j,b}^{uu}(0)-\\
&-Q_{i,d}^{cu}(x)Q_{j,b}^{uc}(0)-Q_{i,d}^{uc}(x)Q_{j,b}^{cu}(0)\big)\big\}\,.
\end{split}
\eeq
From dimensional analysis, the size of this contribution is roughly 
\beq
{\cal H}_{cc}\sim\frac{G_F^2}{16 \pi^2}m_c^2 \lambda_c^d\lambda_c^b 
(\bar s d)_{V-A} (\bar s b)_{V-A}\,,
\eeq
which is comparable to \eqref{DeltaS=2} and needs to be kept. 
Another set of contributions of comparable size coming from double $\Delta S=1$ weak Hamiltonian insertions
has CKM structure $\lambda_c^d\lambda_t^b$. 
The nonlocal contributions
proportional to $\lambda_t^d\lambda_t^b$, on the other hand, are power suppressed, scaling as 
$m_c^2$, compared to the corresponding ones in \eqref{DeltaS=2}, which scale as $m_t^2$. 
These contributions can be safely neglected.

The appearance of nonlocal contributions is similar to the situation for $K^0-\bar K^0$ mixing, 
where the effective Hamiltonian below the charm scale contains the T-product of two 
$\Delta S = 1$ operators mediating $s \to d u \bar u$ transitions, in addition to the local
operator $(\bar s d)_{V-A}(\bar s d)_{V-A}$.
The only difference is that in exclusive $b\to ss \bar d$ decays the charm quark can not
be integrated out because of the large momenta of the light mesons in the final state. 

The dominant nonlocal operators have CKM structure $\lambda_c^d \lambda_t^b$ Eq.~(\ref{tree-T}) and
$\lambda_c^d \lambda_c^b$ Eq.~(\ref{nonlocal-cc}).  These operators contribute to the
physical decay amplitude through rescattering effects with $D\bar D, D\pi, \bar D\pi, \cdots$ 
intermediate states. Their matrix elements are suppressed relative
to those of the top box contribution $\sim C_{tt}$ by $\lambda^{-4} (m_c^2/m_t^2) \simeq 30\%$,
which suggests that the approximation of neglecting $m_c^2$ suppressed 
(but CKM enhanced) nonlocal terms may be a reasonable first attempt. 

We leave a complete calculation of the nonlocal contributions for the future and present only a partial evaluation 
of $b\to ss \bar d$ branching ratios by relating the matrix elements of the local contributions
\eqref{DeltaS=2} to the already measured charmless two body decays using flavor SU(3). 
We note that the nonlocal contributions
were estimated in Ref.~\cite{Fajfer:2000ax} using a hadronic saturation model, and were found to be
suppressed relative to the local contributions. 

For the purpose of the SU(3) relations to be discussed below, it is 
useful to rewrite the effective Hamiltonian \eqref{DeltaS=2} as
\begin{eqnarray}\label{eq:Hi}
{\cal H}_i = \frac{G_F}{\sqrt2} V_{ub}V^*_{ud} \kappa_i O_i\,,\qquad i = ss\bar d, dd\bar s.
\end{eqnarray}
The operators $O_i$ are 
\beq\label{Oi}
O_{ss\bar d} = (\bar s b)_{V-A} (\bar s d)_{V-A}, \qquad
O_{dd\bar s} = (\bar d b)_{V-A} (\bar d s)_{V-A},
\eeq
and the dimensionless coefficients $\kappa_i$ depend only on the CKM factors and
calculable hard QCD coefficients. We have 
\beq
\kappa_{ss\bar d} = \frac{\sqrt2 G_F m_W^2}{(4\pi)^2} \frac{V_{tb} V^*_{ts}}{V_{ub}V^*_{ud}}
( V_{td} V^*_{ts} C_{tt} + V_{cd} V^*_{cs} C_{ct} ),
\eeq
and similarly for $\kappa_{dd\bar s}$. Numerically, the coefficients are (at $\mu=m_b=4.2$ GeV, with CKM elements from \cite{Charles:2004jd}) 
\beq\label{kappa}
\kappa_{ss\bar d} = (6.9 \cdot 10^{-6}) e^{i 51^\circ}\,,\qquad \kappa_{dd\bar s}= (1.5 \cdot 10^{-6}) e^{-i 74^\circ}\,.
\eeq

The SU(3) symmetry relations derived below require also the $C_1+C_2$ combination of Wilson 
coefficients, evaluated at the same scale $\mu=m_b$. 
At leading log order this is given by
\begin{eqnarray}
(C_1 + C_2)(m_b) = \left( \frac{\alpha_s(M_W)}{\alpha_s(m_b)}\right)^{6/23} = 
\eta_2(m_b)=0.85\,.
\end{eqnarray}

\section{SU(3) predictions}
\label{Sec:SU3}

We next show how two body $B$ decay widths for $b\to ss\bar d$ and $b\to dd\bar s$
transitions can be calculated using flavor SU(3). As a first approximation we neglect the
nonlocal charm-top contributions, as justified in the previous section. 
Then the processes $b\to ss\bar d$ and $b\to dd\bar s$ are mediated in the SM
only by the local operators $O_i$ in Eq.~(\ref{Oi}).
Under flavor SU(3) these operators transform as $\overline{\mathbf 15}$ 
\beq\label{Oi:transf}
O_{ss\bar d} = \overline{\mathbf{15}}_{1/2},\qquad O_{dd\bar s} = \overline{\mathbf{15}}_{1},
\eeq
where the subscripts denote the isospin. They belong to the same SU(3) multiplet as the $\overline{\mathbf 15}$ 
in the decomposition of the $b\to du\bar u$ tree operators \cite{Gronau:1998fn}
\beq
\begin{split}
  C_1 &(\bar u b)_{V-A} (\bar d u)_{V-A} + 
C_2 (\bar d b)_{V-A} (\bar u u)_{V-A} = \\
&  \frac12 (C_1+C_2) (- \frac{2}{\sqrt3} \overline{\mathbf{15}}_{3/2}
- \frac{1}{\sqrt6} \overline{\mathbf{15}}_{1/2} + 
\frac{1}{\sqrt2} \overline{\mathbf 3}^{(s)}_{1/2}) 
+ \frac12 (C_1-C_2) ({\mathbf 6}_{1/2} - \overline{\mathbf 3}^{(a)}_{1/2})\,.
\label{DeltaS0}
\end{split}
\eeq
These operators contribute to $\Delta S = 0$ decays such as $B\to \pi\pi$. The explicit expressions for $\overline{\mathbf{15}}$ operators in \eqref{DeltaS0}
are
\begin{eqnarray}
\overline{\mathbf{15}}_{1/2} &=& 
- \frac{1}{2\sqrt6} [(\bar bu)(\bar ud) + (\bar bd)(\bar uu)] +
\frac12 \sqrt{\frac32} [(\bar bs)(\bar sd) + (\bar bd)(\bar ss)] -
\frac{1}{\sqrt6} (\bar bd)(\bar dd), \\
\overline{\mathbf{15}}_{3/2} &=& 
- \frac{1}{\sqrt3} [(\bar bu)(\bar ud) + (\bar bd)(\bar uu)] +
\frac{1}{\sqrt3} (\bar bd)(\bar dd) \,.
\end{eqnarray}

We list the $b\to dd\bar s, ss\bar d$ exclusive decays in Table \ref{table:ampsPP} for $B\to PP$ and in Table \ref{table:ampsPV} for $B\to PV$. The $PP$ final states  transform  under SU(3) as 
$\mathbf{1}, \mathbf{8}, \mathbf{27}$, the operators $O_{ss\bar d}$ and  $O_{dd\bar s}$ are in $\overline{\mathbf{15}}$,  and thus there are only two 
reduced matrix elements,
$\langle \mathbf{8}|\overline{\mathbf{15}}|\mathbf{3}\rangle\,,
\langle \mathbf{27}|\overline{\mathbf{15}}|\mathbf{3}\rangle$.  
These two reduced matrix elements also appear in the predictions for measured 
$\Delta S =0$ decays mediated by the operators in Eq.~(\ref{DeltaS0}).
This means that the $B\to PP$ matrix elements of the operators 
$O_{ss\bar d}$ and $O_{dd\bar s}$ can be expressed in terms of $\Delta S =0$
decay amplitudes such as $A(B^0\to \pi^+\pi^-)$ and others. 
A similar analysis applies to $B\to PV$ decays, where there
are four independent reduced matrix elements of the $\overline{\mathbf{15}}$
operators: 
$\langle \mathbf{8}_S|\overline{\mathbf{15}}|\mathbf{3}\rangle\,,
\langle \mathbf{8}_A|\overline{\mathbf{15}}|\mathbf{3}\rangle\,,
\langle \overline{\mathbf{10}}|\overline{\mathbf{15}}|\mathbf{3}\rangle\,,
\langle \mathbf{27}|\overline{\mathbf{15}}|\mathbf{3}\rangle$. 
These can again be  expressed in terms of physical $B\to PV$ $\Delta S = 0$
amplitudes. We now derive these relations separately for the
$B\to PP$ and $B\to PV$ final states.


\begin{table}
\begin{tabular}{ccc}
\hline
\hline
 Transition & Mode & Amplitude \\ 
\hline
\hline
$b\to ss\bar d$~~ & $B^+ \to K^+K^0$   & $t+c$  \\
                & $B^0 \to K^0 K^0$  & $t+c$   \\
                & $B_s \to K^0 \pi^0$  & $\frac{1}{\sqrt2}(a+e)$   \\
                & $B_s \to K^+ \pi^-$  & $-(a+e)$   \\
                & $B_s \to K^0 \eta_8$  & $\sqrt{\frac23}(t+c+a+e)$   \\
\hline
$b\to dd\bar s$~~ & $B^+ \to \bar K^0 \pi^+$    & $t+c$   \\
                & $B^0 \to \bar K^0 \pi^0$    & $\frac{1}{\sqrt2}(t+c+a+e)$   \\
                & $B^0 \to K^- \pi^+$         & $-(a+e)$   \\
                & $B_s \to \bar K^0 \bar K^0$ & $t+c$   \\
\hline\hline
\end{tabular}
\caption{$B\to PP$ exclusive decays mediated by the $b\to ss\bar d$ and $b\to dd\bar s$
transitions.}
\label{table:ampsPP}
\end{table}

\subsection{$B \to PP$ decays}

We use the formalism of the graphical amplitudes \cite{Gronau:1994rj}, which makes the derivation of 
SU(3) decompositions quite intuitive. The two independent reduced matrix elements of the 
$\overline{\mathbf{15}}$ operator
are given in terms of graphical amplitudes \cite{Grinstein:1996us,Gronau:1998fn} as
\begin{align}
-\frac{\sqrt{10}}{6}\langle {\mathbf 27}|\overline{\mathbf 15}|{\mathbf 3}\rangle &= 
- \frac{1}{C_1+C_2} (T+C)=- \frac{1}{\kappa_{ss\bar d}} (t+c), \label{eq:t+c}\\
\langle {\mathbf 8}|\overline{\mathbf 15}|{\mathbf 3}\rangle &=
- \frac{1}{C_1+C_2} \Big(\frac15 (T+C) + A+E\Big)=- \frac{1}{\kappa_{ss\bar d}} \Big(\frac15 (t+c) + a+e\Big).
\end{align}
This gives two relations between the graphical amplitudes $T$ (tree), $C$ (color-suppressed tree), 
$A$ (annihilation), $E$  (exchange) in the $\Delta S = 0$ modes (the expression for $B\to PP$ decays can be 
found in \cite{Gronau:1994rj}) and the corresponding graphical amplitudes $t,c,a,e$ in $b\to ss\bar d$ transitions 
(the decay amplitudes for $B\to PP$ modes in terms of these are collected in Table~\ref{table:ampsPP}). 
Equivalent relations apply between $\Delta S=0$ and $b\to dd\bar s$ decay amplitudes.

The most useful for our purposes is the relation \eqref{eq:t+c}. This gives the following prediction for the exclusive $b\to ss\bar d$
decays
\begin{eqnarray}\label{27}
A(B^+ \to K^+ K^0 ) = A(B^0 \to K^0 K^0) = \frac{\kappa_{ss\bar d}}{C_1+C_2}
\sqrt2 A(B^+ \to \pi^+\pi^0)\,,
\end{eqnarray}
and similarly for the $b\to dd\bar s$  decay
\begin{eqnarray}
A(B^+ \to \bar K^0 \pi^+ ) =A(B_s \to \bar K^0 \bar K^0)= \frac{\kappa_{dd\bar s}}{C_1+C_2}
\sqrt2 A(B^+ \to \pi^+\pi^0)\,.
\end{eqnarray}
Neglecting the $1/m_b$ suppressed amplitudes $e,a$ one also has
\beq
{\frac{\sqrt3}{2}}A(B_s \to K^0 \eta_8 )= \frac{\kappa_{ss\bar d}}{\kappa_{dd\bar s}}A(B^0 \to \bar K^0 \pi^0)\simeq\frac{\kappa_{ss\bar d}}{C_1+C_2}A(B^+ \to \pi^+\pi^0)\,.
\eeq
The remaining amplitudes in Table~\ref{table:ampsPP} are proportional to $e,a$. 
They are $1/m_b$ suppressed, therefore we do not consider them further. 

The same SU(3) relations hold also for the decays into two vector mesons, $B \to V_\lambda V_\lambda$, 
separately for each helicity amplitude
$\lambda = 0,\pm$. For example, the analog of Eq.~(\ref{27}) is
\begin{eqnarray}
A(B^+\to K^{*+}_\lambda K^{*0}_\lambda) = A(B^0 \to K^{*0}_\lambda K^{*0}_\lambda) =
\frac{\kappa_{ss\bar d}}{C_1+C_2} \sqrt2 A(B^+\to \rho^+_\lambda \rho^0_\lambda).
\end{eqnarray}
As a consequence the $b\to ss\bar d$ and $b\to dd\bar s$ $B\to VV$ decays are longitudinally 
polarized in the same way as the $B^+\to \rho^+\rho^0$ decay.

\subsection{$B\to PV$ decays}

\begin{table}
\begin{tabular}{ccc}
\hline\hline
 Transition & Mode & Amplitude \\ 
\hline
\hline
$b\to ss\bar d$~ & $B^+ \to K^{*+}K^0$ & $t_V + c_V$ \\
                & $B^+ \to K^{+}K^{*0}$ & $t_P + c_P$     \\
                & $B^0 \to K^{*0} K^0$  & $t_P+t_V+c_P+c_V$  \\
                & $B_s \to K^{*0}\pi^0$  & $\frac{1}{\sqrt2}(a_P+e_P)$  \\
                & $B_s \to K^0\rho^0$    & $\frac{1}{\sqrt2}(a_V+e_V)$  \\
                & $B_s \to K^{*0}\eta_8$ & $\sqrt{\frac23}(t_P+c_P+a_V+e_V)$  \\
                & $B_s \to K^0\phi_8$    & $\sqrt{\frac23}(t_V+c_V+a_P+e_P)$  \\
                & $B_s \to K^{*+}\pi^-$  & $-(a_P+e_P)$  \\
                & $B_s \to K^+\rho^-$    & $-(a_V+e_V)$  \\
\hline
$b\to dd\bar s$~ & $B^+ \to \bar K^{*0} \pi^+$ & $t_P + c_P$ \\
                & $B^+ \to \bar K^0\rho^+$    & $t_V + c_V$ \\
                & $B^0 \to \bar K^{*0} \pi^0$ & $\frac{1}{\sqrt2}(t_P+c_P+a_V+e_V)$  \\
                & $B^0 \to \bar K^0\rho^0$    & $\frac{1}{\sqrt2}(t_V+c_V+a_P+e_P)$ \\
                & $B^0 \to K^{*-}\pi^+$       & $-(a_V+e_V)$  \\
                & $B^0 \to K^-\rho^+$         & $-(a_P+e_P)$  \\
                & $B_s \to \bar K^{*0}\bar K^0$ & $t_V+t_P+c_V+c_P$  \\
\hline\hline
\end{tabular}
\caption{$B\to PV$ exclusive decays mediated by the $b\to ss\bar d$ and $b\to dd\bar s$
transitions.}
\label{table:ampsPV}
\end{table}

Table~\ref{table:ampsPV} lists the decomposition of $B\to PV$ decays in terms of graphical amplitudes.
The subscripts $P,V$ on $t,c$ identify the final state meson that contains the spectator quark,
while the subscripts on $a,e$ denote the final state meson containing the $q_3$ quark from $\bar b\to \bar q_1\bar q_2q_3$ (here the spectator participates in the weak interaction) \cite{Dighe:1997wj,Gronau:2000az}.

We have $T_{P,V}+C_{P,V}\propto \langle \overline{\mathbf{10}}|\overline{\mathbf{15}}|\mathbf{3}\rangle\pm
\langle \mathbf{27}|\overline{\mathbf{15}}|\mathbf{3}\rangle$.
The analogs of the relation (\ref{eq:t+c}) are then
\beq\label{eqPV}
 t_P + c_P = \frac{\kappa_{ss\bar d}}{C_1+C_2} (T_P+C_P),\qquad t_V + c_V = \frac{\kappa_{ss\bar d}}{C_1+C_2} (T_V+C_V)\,,
\eeq
where the graphical amplitudes on the right-hand side are for $\Delta S = 0$ decays. The expansion of the corresponding 
decay amplitudes in terms of graphical amplitudes can be found in  Refs.~\cite{Dighe:1997wj,Gronau:2000az}. Combining
them with expansions in Table~\ref{table:ampsPV} gives the SU(3) relations
for the $t_i+c_i$ exclusive $b\to ss\bar d$ decay amplitudes (for $\Delta S=0$ amplitude we only denote the 
final state)
\begin{align}
\begin{split}
& A(B^+ \to K^{*+} K^0) = \frac{\kappa_{ss\bar d}}{C_1 + C_2}
\Big[ -\big(A_{\rho^+\pi^-} - A_{\rho^-\pi^+}\big) - \sqrt2 A_{\rho^0\pi^+}\\
&  \quad + \big(A_{K^{*0} \bar K^0} - A_{\bar K^{*0} K^0}\Big) - 
      \big(A_{K^{*-}K^+} - A_{K^{*+} K^-}\big) + 
      \big(A_{\bar K^{*0} K^+} - A_{K^{*+} \bar K^0}\big) \big],
\end{split}      
\\
\begin{split}
& A(B^+ \to K^{+} K^{*0}) = \frac{\kappa_{ss\bar d}}{C_1 + C_2}
\Big[ \big(A_{\rho^+\pi^-} - A_{\rho^-\pi^+}\big) - \sqrt2 A_{\rho^+\pi^0}\\
&  \quad - \big(A_{K^{*0} \bar K^0} - A_{\bar K^{*0} K^0}\big) + 
      \big(A_{K^{*-}K^+} - A_{K^{*+} K^-}\big) - 
      \big(A_{\bar K^{*0} K^+} - A_{K^{*+} \bar K^0}\big) \Big], 
\end{split}\\
& A(B^0 \to K^{*0} K^0) = -\frac{\kappa_{ss\bar d}}{C_1 + C_2}
    \sqrt2 \big(A_{\rho^+\pi^0} + A_{\rho^0\pi^+}\big)\,.
\end{align}
The $B_s$ decay amplitudes containing $t_i+c_i$ are given in terms of the above $b\to ss\bar d$ amplitudes
\begin{align}
& A(B_s \to K^{*0} \eta_8) = \sqrt{\frac23} \big[A(B^+ \to K^+ \bar K^{*0}) - A(B_s \to K^+\rho^-)\big],\\
& A(B_s \to K^{0} \phi_8) = \sqrt{\frac23} \big[A(B^+ \to K^{*+} K^{0}) - A(B_s \to K^{*+}\pi^-)\big]\,,
\end{align}
where the $1/m_b$ suppressed pure annihilation and exchange decay amplitudes are
\begin{align}
A(B_s \to K^{*+} \pi^-) \!=\!  -\sqrt2 A(B_s\to K^{*0} \pi^0) \!=\! - \frac{\kappa_{ss\bar d}}{C_1+C_2} 
[A_{\bar K^{*0} K^+} - A_{K^{*-} K^+} - A_{\bar K^{*0} K^0}]\,, \\
A(B_s \to K^{+} \rho^-) \!=\!  -\sqrt2 A(B_s \to K^0 \rho^0) \!=\! - \frac{\kappa_{ssd}}{C_1+C_2}
   [A_{K^{*+} \bar K^0} - A_{K^{*+} K^-} - A_{K^{*0} \bar K^0}]\,.
\end{align}
The relations for the $b\to dd\bar s$ transitions are derived in an analogous way, giving for the 
$t_i+c_i$ amplitudes
\begin{align}
\begin{split}
&A(B^+\to \bar K^{*0} \pi^+) = \frac{\kappa_{dd\bar s}}{C_1+C_2}
\Big[ \big(A_{\rho^+\pi^-} - A_{\rho^-\pi^+}) - \sqrt2 A_{\rho^+\pi^0}\\
& \quad - \big(A_{K^{*0} \bar K^0} - A_{\bar K^{*0} K^0}\big) + 
      \big(A_{K^{*-}K^+} - A_{K^{*+} K^-}\big) - 
      \big(A_{\bar K^{*0} K^+} - A_{K^{*+} \bar K^0}\big) \Big], 
\end{split}     
\\
\begin{split}
& A(B^+ \to \bar K^{0} \rho^+) = \frac{\kappa_{dd\bar s}}{C_1 + C_2}
\Big[ -\big(A_{\rho^+\pi^-} - A_{\rho^-\pi^+}\big) - \sqrt2 A_{\rho^0\pi^+}\\
&  \quad + \big(A_{K^{*0} \bar K^0} - A_{\bar K^{*0} K^0}\big) - 
      \big(A_{K^{*-}K^+} - A_{K^{*+} K^-}\big) + 
      \big(A_{\bar K^{*0} K^+} - A_{K^{*+} \bar K^0}\big) \Big],
\end{split}
\end{align}
and
\begin{eqnarray}
& &\sqrt2 A(B^0 \to \bar K^{*0} \pi^0) = A(B^+ \to \bar K^{*0} \pi^+) - 
A(B^0 \to K^{*-} \pi^+),\\
& &\sqrt2 A(B^0 \to \bar K^{0} \rho^0) = A(B^+ \to \bar K^{0} \rho^+) - 
A(B^0 \to K^{-} \rho^+)\,.
\end{eqnarray}
The $1/m_b$ suppressed pure annihilation and exchange amplitudes are
\begin{align}
& A(B^0 \to K^{*-} \pi^+) = \frac{\kappa_{dd\bar s}}{C_1 + C_2}
\big[  -A_{K^{*+} \bar K^0} + A_{K^{*+} K^-} + A_{K^{*0} \bar K^0} \big], \\
& A(B^0 \to K^{-} \rho^+) = \frac{\kappa_{dd\bar s}}{C_1 + C_2}
\big[  -A_{\bar K^{*0} K^+} + A_{K^{*-} K^+} + A_{\bar K^{*0} K^0} \big] \,.
\end{align}
The  remaining $B_s$ mode is given by
\begin{eqnarray}
A(B_s \to \bar K^{*0} \bar K^0) = 
- \frac{\kappa_{dd\bar s}}{C_1+C_2} \sqrt2 [A_{\rho^+\pi^0} + A_{\rho^0\pi^+}].
\end{eqnarray}

These SU(3) relations will be used in the next Section to predict branching 
fractions of exclusive $b\to ss\bar d$ and $b\to dd\bar s$ decays in the SM. The measured branching 
fractions of $\Delta S =0$ modes are then the inputs in the predictions and are collected in Table \ref{table-1}. 
We only quote results for those decays that are not $1/m_b$ suppressed.

\section{SM predictions from the SU(3) relations}\label{SMpredictions}

Experimentally one will be able to search for NP effects in the following $b\to ss \bar d$ decays $\bar B^0\to \bar K^{0*}\bar K^{0*}$, $B^-\to K^- \bar K^{0*}$, $\bar B_s^0\to \phi \bar K^{0*}$. The flavor of $\bar K^{0*}$ is tagged using the decay $K^{0(*)}\to K^+\pi^-$. The same decays
with $\bar K^0$ instead of $\bar K^{0*}$, on the other hand, cannot be used to probe $b\to ss \bar d$ transitions. The $K^0$ mixes with $\bar K^0$ so that mass eigenstates $K_{S,L}$ are observed in the experiment. The "wrong kaon" decays listed above 
are thus only a subleading contribution in the SM rate. For easier comparison with previous calculations in the literature we will still
quote results for $\bar B^0\to \bar K^{0}\bar K^{0}$, $\dots$, "branching ratios", knowing that these are unobservable in practice.
Similar comments apply to $b\to dd\bar s$ transitions, where NP effects can be probed 
in $\bar B^0\to \pi^0 K^{0*}, \rho^0 K^{0*}$, $B^-\to \pi^- K^{0*}, \rho^- K^{0*}$ 
and $\bar B_s^0\to K^{0*} K^{0*}$ decays, again using flavor tagged $K^{0*}$ decays.

We derive next numerical predictions for the branching fractions of the
exclusive $b\to ss\bar d, dd\bar s$ modes. 
The branching fraction of a given mode $B_q \to M_1 M_2$ is given by
\begin{eqnarray}
{\cal B}(B_q \to M_1 M_2) = 
\tau_{B_q} |A(B_q \to M_1 M_2)|^2 \frac{|\vec p\,|}{8\pi m_{B_q}^2}
\end{eqnarray}
To predict $b\to ss\bar d, dd\bar s$ decay amplitudes, $A(B_q \to M_1 M_2)$, we use the SU(3) relations derived
in Sec.~\ref{Sec:SU3} which relate them   to the amplitudes of the already  measured $B^+\to \pi^+\pi^0, \rho^+\rho^0$, and $B\to \rho \pi$ decays. The results are collected in Tables \ref{table:results} and \ref{table:resultsdds}. 
As mentioned, we do not present results for the branching ratios of the $1/m_b$ suppressed 
annihilation modes. 

In the calculation of $B\to PV$ branching ratios we neglect the contributions of the
small penguin dominated $B\to \bar K^* K, K^* \bar K$ decays in the SU(3) relations 
(with experimental upper bounds supporting this approximation). Furthermore, the application 
of the SU(3) relations requires
that we know also the relative phases of the $B\to \rho\pi$ amplitudes.
These phases are small, and can be neglected to a good approximation. 
This can be verified using the isospin pentagon relation
\begin{eqnarray}
A(\rho^+\pi^0) + A(\rho^0\pi^+) = 
\frac{1}{\sqrt{2}}(A(\rho^+\pi^-) + A(\rho^-\pi^+))
+ \sqrt{2} A(\rho^0\pi^0)\,.
\end{eqnarray}
Neglecting the relative phases, and using data from Table~\ref{table-1}, the left-hand side of this equality is
$6.25\pm 0.29$ (in units of $\sqrt{{\cal B} \cdot 10^6}$), which compares well with the right-hand side of
$6.91\pm 0.34$. This justifies the assumption made of neglecting the
relative phases of the $B \to \rho\pi$ amplitudes.

To factor out the dependence on CKM elements, we also quote the predictions for 
$B\to PP, PV,VV$ modes in a common form as 
\begin{eqnarray}\label{BrPrediction}
{\cal B}(B\to X_i) = \frac{|\kappa_{ssd}|^2}{(C_1+C_2)^2} c_i,
\end{eqnarray}
where $c_i$ are coefficients specific to each final state calculated using the
SU(3) relations and measured $\Delta S =0$ branching fractions.
In the predictions we used the branching fractions for the $\Delta S = 0$ modes listed in Table
\ref{table-1}. We use 
$\tau(B^+)/\tau(B^0)=1.071 \pm 0.009$ and $\tau(B_s^0)/\tau(B^0)=0.965 \pm 0.017$ \cite{HFAG}. 

\begin{table}
\begin{tabular}{cccc}
\hline\hline
Mode &    $c_i [\times 10^{-6}]$ &  ${\cal B}_{\rm SM}$ & Literature \\ 
\hline\hline
$B^+ \to K^+K^0$~& ~$11.0\pm 0.8$~  & ~$(0.7 \pm 0.1) \cdot 10^{-15}$~ & ~$2.5\times 10^{-14}$~ \\
$B^0 \to K^0 K^0$ & $10.2\pm 0.7$  & $(0.7 \pm 0.1) \cdot 10^{-15}$ & $-$  \\
\hline
$B^+ \to K^{*+}K^0$ & $29.3\pm 4.3$  & $(1.9 \pm 0.3) \cdot 10^{-15}$ & $1.7\times 10^{-14}$ \\
$B^+ \to K^{+}K^{*0}$ & $11.3\pm 3.0$  & $(0.7 \pm 0.2) \cdot 10^{-15}$ & $6.5\times 10^{-14}$ \\
$B^0 \to K^{*0}K^0$ & $71.5\pm 6.2$  & $(4.7 \pm 0.4) \cdot 10^{-15}$ & $-$ \\
\hline
$B^+ \to K^{*+}K^{*0}$  & $47.2\pm 3.7$  & $(3.1 \pm 0.2) \cdot 10^{-15}$ & $6.8\times 10^{-14}$ \\
$B^0 \to K^{*0} K^{*0}$ & $43.9\pm 3.5$  & $(2.9 \pm 0.2) \cdot 10^{-15}$ & $-$  \\
\hline\hline
\end{tabular}
\caption{SU(3) predictions for the branching fractions of the $b\to ss\bar d$ modes in the SM. 
The last column shows the predictions from a previous calculation \cite{Fajfer:2000ny}.}
\label{table:results}
\end{table}

\begin{table}
\begin{tabular}{ccc}
\hline\hline
Mode &    $c_i [\times 10^{-6}]$ &  ${\cal B}_{\rm SM} $  \\ 
\hline\hline
~$B^+ \to \bar K^0\pi^+$~ & ~$11.1\pm 0.8$~  & ~$(35 \pm 2) \times 10^{-18}$~ \\
$B_s \to \bar K^0 \bar K^0$ & $9.7\pm 0.7$  & $(30 \pm 2) \times 10^{-18} $   \\
\hline
$B^+ \to \bar K^{*0}\pi^+$ & $11.4\pm 2.9$  & $(36 \pm 9)  \times 10^{-18}$  \\
$B^+ \to \bar K^{0}\rho^{+}$ & $29.5\pm 4.3$  & $(92 \pm 13) \times 10^{-18}$  \\
$B^0 \to \bar K^{*0}\pi^0$ & $5.3\pm 1.4$  & $(17 \pm 4) \times 10^{-18} $  \\
$B^0 \to \bar K^0\rho^0$   &  $13.7\pm 2.0$  &   $(43 \pm 6) \times 10^{-18} $          \\
$B_s \to \bar K^{*0} \bar K^0$ &  $69.1\pm 6.0$  &  $(215 \pm 19) \times 10^{-18} $       \\
\hline
$B^+ \to \bar K^{*0}\rho^+$ & $47.6\pm 3.8$  & $(148 \pm 12) \times 10^{-18}$  \\
$B_s \to \bar K^{*0} \bar K^{*0}$ & $41.7\pm 3.3$  & $(130 \pm 10) \times 10^{-18} $  \\
\hline\hline
\end{tabular}
\caption{SU(3) predictions for the branching fractions of the $b\to dd\bar s$ modes in the SM. }
\label{table:resultsdds}
\end{table}

Both Belle \cite{Garmash:2003er} and BABAR \cite{Aubert:2008rr} 
collaborations presented the results of a search for these modes
and report the $90\%$ C.L. upper bounds (BABAR bounds are in square brackets)
\begin{align}
\begin{split}
b\to dd\bar s:~~& {\cal B}(B^+ \to K^-\pi^+\pi^+) < 45.0 \;[9.5] \times 10^{-7}; \quad\text{BELLE\;[BABAR]},
\end{split}\\
\label{boundK+Kst0}
\begin{split}
b \to ss\bar d:~~& {\cal B}(B^+ \to K^+ K^+\pi^-) < 24.0 \;[9.5] \times 10^{-7};\quad\text{BELLE\;[BABAR]}, 
  \end{split}\\
               & {\cal B}(B^0 \to K^0 K^+\pi^-) < 180 \times 10^{-7}; \quad \text{BELLE}.
\end{align}
The quasi two--body decay $B^+\to \bar K^{0*} \pi^+$ is part of the $B^+ \to K^-\pi^+\pi^+$ three body decay, 
$B^+\to K^+K^{*0}$ is part of $B^+ \to K^+ K^+\pi^-$, while $B^0\to K^0 K^{*0}$ is part of $B^0 \to K^0 K^+\pi^-$.
The bounds on three body decays thus imply bound on two-body decays. These are 8 orders of magnitude or more above the estimates for the SM signal, but the situation could
improve at a future super-B factory \cite{Browder:2008em} or at LHCb. Note that $B^0 \to K^0 K^+\pi^-$ is observed in $K_S K^+\pi^-$ final states which also receives contributions from  $b\to d$ penguin decay $B^0\to \bar K^0 K^{*0}$ and from annihilation decay $B^0\to K^+K^-$. It thus cannot be used as a null probe of NP. 

\begin{table}
\newcommand{\cerr}[3]   {\mbox{${{#1}^{+ #2}_{- #3}}$}}
\label{table-1}
\begin{tabular}{lllclllc}
\hline\hline
\multicolumn{2}{c}{Mode} &  ${\cal B}(\times 10^{-6})$ & \multicolumn{2}{c}{Mode} & ${\cal B}(\times 10^{-6})$ \\
\hline
$B^+\to$
&$\pi^0\pi^+$ & ~~$\cerr{5.59}{0.41}{0.40}$~~ &
$B^0 \to$ & $\rho^{\pm} \pi^\mp$ & ~~$23.0 \pm 2.3$~~ 
\\
&$\rho^0\pi^+$ & ~~$\cerr{8.7}{1.0}{1.1}$~~ &
& $\rho^+\pi^-$ & ~~$15.4\pm 1.8^{a}$ 
\\
&$\rho^+\pi^0$ & ~~$\cerr{10.9}{1.4}{1.5}$~~ &
& $\rho^-\pi^+$ & ~~$7.2\pm 1.1^{b}$~~
\\
&$\rho^+\rho^0$ & ~~$2.40 \pm 0.19$~~ &
&$\rho^0\pi^0$ & ~~$2.0 \pm 0.5$~~ \\
\hline\hline
\end{tabular}
\caption{\footnotesize{Branching ratios for $B\to \pi\pi,\rho\pi,\rho\rho$ decays, from Ref.~\cite{HFAG} apart from: 
$a$) the average of $15.5\pm 3.4$~\cite{Kusaka:2007mj} and $15.3 \pm 2.2$~\cite{Aubert:2006fg}, and 
$b$) the average of $7.1\pm 1.9$~\cite{Kusaka:2007mj} and $7.3\pm 1.4$~\cite{Aubert:2006fg}.
}}
\end{table}

\section{$b\to ss\bar d$ and $b\to dd\bar s$  transitions in the presence of NP}
\label{NPsection}
Next we consider the $b\to ss\bar d$ and $b\to dd\bar s$ decays in the presence 
of generic NP. The most general local NP hamiltonian 
mediating the $b\to ss\bar d$ and $b\to dd\bar s$ transitions was given in Eq.~(\ref{HeffNP}).
In this section we will assume that NP matches onto the local operator $Q_1$
in Eq.~(\ref{HeffNP}) with SM chirality $(V-A)\times (V-A)$. 
This is true for a large class of 
NP models, such as the two-Higgs doublet model with small $\tan\beta$, 
or the constrained MSSM \cite{Fajfer:2006av}. Effects of NP that matches
to other chiral structures will be given in the next section.\footnote{There is also 
the possibility that NP contributes through the $\Delta S = 1$ hamiltonians
appearing in the nonlocal term in Eq.~(\ref{HeffSM}), for instance through a
$(\bar s b)(\bar c c)$ term. We do not pursue this possibility further. } 
To simplify the notation we focus on $b\to ss\bar d$ transitions --- the expressions for $b\to dd\bar s$ 
can be obtained through a simple $s\leftrightarrow d$ exchange --- but show numerical results for both 
types of decays.

We consider three representative cases of NP: i) the exchange of NP fields that carry a conserved charge, 
where large effects are possible as explained in the Introduction, 
ii) minimally flavor violating (MFV) new physics with small $\tan\beta$ 
\cite{D'Ambrosio:2002ex}, NMFV \cite{Agashe:2005hk} and
iii) general flavor violation with a $\sim 10^3$ TeV scale suppression. For this analysis it is useful to 
rewrite the $b\to ss\bar d$ SM effective Hamiltonian \eqref{eq:Hi} as
\begin{eqnarray}
{\cal H}_{ss\bar d} = \frac{G_F}{\sqrt2} V_{ub}V^*_{ud} \kappa_{ss\bar d} O_{ss\bar d}=\frac{1}{\Lambda_0^2} e^{-i\gamma}\kappa_{ss\bar d} Q_1\,,
\end{eqnarray}
where $\Lambda_0=2^{1/4}/(2 \sqrt{G_F|V_{ub}V_{ud}|})=2.98$ TeV and $Q_1$ is defined in \eqref{Qops} (the flavor dependence of $Q_1$ is not shown). The NP Hamiltonian for $b\to ss\bar d$ is 
\beq\label{NP-hamilt}
{\cal H}_{ss\bar d}^{\rm NP}=\frac{c_1}{\Lambda_{\rm NP}^2} \eta_2 Q_1,
\eeq
where the  Wilson coefficient $c_1$ contains possible extra flavor hierarchy in the new physics transitions, 
and $\Lambda_{\rm NP}$ is the scale of NP. The hard QCD correction to the $c_1$ coefficient describing the
RG running from the weak scale $m_W$ 
to $m_b$ has been explicitly factored out, $\eta_2(m_b)=0.85$. It is now very easy to obtain the branching 
ratio in the presence of NP from the SM predictions, 
\beq
{\cal B}_{\rm NP}(B\to f)=
\Big|\frac{ \Lambda_0^2 }{\kappa_{ss\bar d}}\frac{c_1\eta_2}{\Lambda_{\rm NP}^2}\Big|^2 {\cal B}_{\rm SM}(B\to f),
\eeq
and similarly for $b\to dd\bar s$ decays.

\subsection{NP with conserved charge}

As discussed in the introduction it is possible to have large NP effects in $b\to ss\bar d$ decays, if the transition is mediated by 
NP fields that carry a total conserved charge, and if in addition there  exists a hierarchy in the couplings. In this case we 
have for the Wilson coefficient in the NP Hamiltonian \eqref{NP-hamilt} (cf. Eq. \eqref{NPX})
\beq
\frac{c_1}{\Lambda_{\rm NP}^2}=\frac{1}{M_X^2}\big(g_{b\to s} g_{s\to d}^*+g_{d\to s}g_{s\to b}^*\big).
\eeq
From $K-\bar K$ and $B_s-\bar B_s$ mixing we have the bounds, Eq. \eqref{bounds},
\beq\label{g-bounds}
\frac{|g_{ d\to s} g_{s\to d}^*|^{1/2}}{M_X}<\frac{1}{10^3 \text{~TeV}},\qquad \frac{|g_{b\to s}g_{s\to b}^*|^{1/2}}{M_X}<\frac{1}{30\text{~TeV}}. 
\eeq
These bounds are trivially satisfied, if for instance $g_{s\to d}=g_{b\to s}=0$, since then no mixing 
contributions are induced. The $b\to ss\bar d$ transitions, on the other hand, can still be large, 
if $g_{d\to s}$ and $g_{s\to b}$ are nonzero. Taking $g_{d\to s}=g_{s\to b}=1$,
the BABAR experimental bound on ${\cal B}(B^+\to K^+K^{*0})$, Eq. \eqref{boundK+Kst0}, gives $M_X>5.0$ TeV\,.
A similar bound $M_X>5.0$ TeV is found for
$b\to dd\bar s$ from the BABAR bound on $B^+\to \bar K^{*0} \pi^+$ branching ratio. 
The resulting predictions for $b\to ss\bar d$ and $b\to dd\bar s$ branching ratios are  of 
$\mathcal{O}(10^{-6})$ as shown in Table~\ref{table:resultsNP} and may well be probed at Belle~II and LHCb.

A more generic situation may be that only one of the $g_i$ couplings is accidentally small. 
Unlike in the previous example, we choose $M_X$ such that we
do not saturate the present experimental bounds on $b\to ss\bar d$. 
As an illustration let us take $g_{s\to d}=0$ and all the other 
couplings to be equal to 1. In this case the $K-\bar K$ mixing bound in \eqref{g-bounds} is trivially 
satisfied, while $B_s-\bar B_s$ mixing implies that $M_X> 30$ TeV. The $b\to ss\bar d$ branching ratios are 
\beq
{\cal B}(\big\{B^+\to K^+K^{*0},B^+\to K^{*+}K^{*0},B^0\to K^{*0}K^{*0}\big\})=\big\{1.1,4.6,4.3\big\}\times 10^{-9} \Big(\frac{30\text{TeV}}{M_X}\Big)^4.
\eeq
For $b\to dd\bar s$ transitions the same choice for the values of coupling, $g_{s\to d}=0$  and all the other $g_i=1$, sets $M_X> 210$ TeV due to the  present absence of NP effects in $B_d-\bar B_d$ mixing. This gives
\beq
{\cal B}(\big\{B^+\to \bar K^{*0}\pi^+,B^+\to \bar K^{*0}\rho^+,B_s^0\to \bar K^{*0}\bar K^{*0}\big\})=
\big\{0.5,1.9,1.7\big\}\times 10^{-12} \Big(\frac{210\text{TeV}}{M_X}\Big)^4.
\eeq
Finally, we mention that, if $b\to ss\bar d$ or $b\to dd\bar s$ modes are observed in the near future, this would imply nontrivial
exclusions on the parameter space of the models. In particular models with $g_{s\to b}\sim g_{s\to d}$ and/or $g_{b\to s}\sim g_{d\to s}$ would be excluded as discussed in Appendix \ref{app:bounds}.

\begin{table}
\begin{tabular}{ccccc}
\hline\hline
Mode $(ss\bar d)$ &  ${\cal B}_{\rm X} [\times (\frac{5.0\text{ TeV}}{M_X})^4]$ &  ${\cal B}_{\rm NMFV} [\times (\frac{173\text{ TeV}}{\Lambda_{ssd}})^4]$ & ${\cal B}_{\rm gen.}$ & ${\cal B}_{\rm SM} $\\
 \hline
$B^+ \to K^+K^{*0}$~    & ~$1.4 \times 10^{-6}$~ & $1.0 \times 10^{-12}$ &$0.3 \times 10^{-14}$ & $(0.7\pm0.2)\times 10^{-15}$\\
$B^+ \to K^{*+} K^{*0}$ & $6.0 \times 10^{-6}$   & $4.2 \times 10^{-12}$ & $1.4 \times 10^{-14}$& $(3.1 \pm0.2)\times 10^{-15}$\\
$B^0 \to K^{*0}K^{*0}$  & $5.5 \times 10^{-6}$   & $3.9 \times 10^{-12}$ & $1.3 \times 10^{-14}$& $(2.9\pm0.2) \times 10^{-15}$\\
\hline\hline
Mode $(dd\bar s)$ &${\cal B}_{\rm X}[\times (\frac{5.0\text{ TeV}}{M_X})^4]$    & ${\cal B}_{\rm NMFV}[\times (\frac{458\text{ TeV}}{\Lambda_{dds}})^4]$ & ${\cal B}_{\rm gen.}$ & ${\cal B}_{\rm SM} $\\
\hline
$B^+\to \bar K^{*0}\pi^+$~  &  ~$1.4\times 10^{-6}$~   & $2.0\times 10^{-14}$& $1.3\times 10^{-15}$ & $(3.6\pm0.9)\times 10^{-17}$\\
$B^+\to \bar K^{*0}\rho^+$  &   $6.0\times 10^{-6}$     & $8.5\times 10^{-14}$& $5.4\times 10^{-15}$ & $(14.8\pm1.2)\times 10^{-17}$\\
$B_s\to \bar K^{*0} \bar K^{*0}$ & $5.3\times 10^{-6}$ & $7.5\times 10^{-14}$& $4.7\times 10^{-15}$  & $(13\pm 1)\times 10^{-17}$\\
\hline\hline
\end{tabular}
\caption{Predictions for the branching fractions of the $b\to ss\bar d, dd\bar s$ modes in the presence
of NP carrying conserved charge (${\cal B}_{\rm X}$), with NMFV flavor structure (${\cal B}_{\rm NMFV}$), 
and general flavor violation at high scale (${\cal B}_{\rm gen.}$), in all cases assuming dominance of the 
SM operator $(\bar s d)_{V-A} (\bar s b)_{V-A}$ (see also text for details). The branching fractions
${\cal B}_{\rm gen}$ include (maximally constructive) interference with the SM amplitude, and were
obtained using $\Lambda_{ssd,dds} = 10^3$ TeV.
Only modes which do not contain $K_{S,L}$ are shown. }
\label{table:resultsNP}
\end{table}

\subsection{NP with MFV and NMFV structures}
Both MFV \cite{D'Ambrosio:2002ex} and NMFV \cite{Agashe:2005hk} fall in the class of new physics models where the 
$b\to ss\bar d$ suppression scale $\Lambda_{ss\bar d}$ is the geometric average of the NP scales 
in $K-\bar K$ and $B_s-\bar B_s$ mixing \eqref{bounds}, $\Lambda_{ss\bar d}\sim \sqrt{\Lambda_{sd}\Lambda_{bs}}\gtrsim 173$ TeV. 
In this paper we will restrict ourselves to MFV with small $\tan\beta$, where the $\Delta F=2$ processes
are mediated by a single operator with $(V-A)\times (V-A)$ structure \cite{D'Ambrosio:2002ex}.
This implies that the $K-\bar K$ and $B_s-\bar B_s$ mixing operators are 
\beq\label{MFVmixing}
\frac{V_{ts}^2 V_{td}^{*2}}{\Lambda_{\rm MFV}^2} (\bar s_L\gamma_\mu d_L)^2\equiv \frac{1}{\Lambda_{s d}^2} (\bar s_L\gamma_\mu d_L)^2, \quad \frac{V_{tb}^2 V_{ts}^{*2}}{\Lambda_{\rm MFV}^2} (\bar b_L\gamma_\mu s_L)^2\equiv \frac{1}{\Lambda_{bs}^2} (\bar b_L\gamma_\mu s_L)^2, 
\eeq
and the $b\to ss\bar d$ local operator is
\beq\label{MFVdecay}
\frac{1}{\Lambda_{\rm MFV}^2} V_{tb} V_{ts}^{*}V_{td} V_{ts}^{*}(\bar b_L\gamma_\mu s_L)(\bar d_L\gamma^\mu s_L)\equiv \frac{1}{\Lambda_{ss\bar d}^2} (\bar b_L\gamma_\mu s_L)(\bar d_L\gamma^\mu s_L), 
\eeq
all of which depend only on one unknown parameter, the MFV scale $\Lambda_{\rm MFV}$. From a global fit the 
UTfit collaboration finds $\Lambda_{\rm MFV}>5.5$ TeV \cite{Bona:2007vi}. We have also defined the suppression 
scales $\Lambda_{sd}$, $\Lambda_{bs}$, $\Lambda_{ss\bar d}$ that include the hierarchy of the NP induced flavor 
changing couplings, which in the MFV case are just the 
appropriate CKM matrix elements. They are related as stated above 
$\Lambda_{ss\bar d}= \sqrt{\Lambda_{sd}\Lambda_{bs}}$. 

In NMFV the operators in \eqref{MFVmixing} and \eqref{MFVdecay} are still parameterically suppressed by 
the CKM matrix elements, but the strict correlation between the Wilson coefficients is lost -- they are 
multiplied by $O(1)$ complex coefficients.
We then have approximately $\Lambda_{ss\bar d}\sim \sqrt{\Lambda_{sd}\Lambda_{bs}}$. 
Using the bounds from \eqref{bounds}, $\Lambda_{ss\bar d}\gtrsim 173$ TeV, which gives $b\to ss\bar d$ 
branching ratios of $\mathcal{O}(10^{-12})$, Table~\ref{table:resultsNP}. Similarly we have 
$\Lambda_{dd\bar s}\sim \sqrt{\Lambda_{sd}\Lambda_{bd}}\gtrsim 458$ TeV, giving $b\to dd\bar s$ branching 
ratios of 
$\mathcal{O}(10^{-13})$. In MFV the predicted branching ratios are much smaller, of the order of 
the SM branching ratios in Table~\ref{table:resultsNP}. The reason for the difference between NMFV and MFV is 
that in MFV NP the Wilson coefficient generating $K-\bar K$ mixing carries a weak phase 
(the same phase as it does in the SM), while in NMFV the NP contribution can be real.

\subsection{General flavor violation with a high scale}
As a final example we consider the case where NP is at the mass scale probed by $K-\bar K$ mixing, 
$\Lambda\sim 10^{3}$ TeV, and assume that flavor violating couplings are all of ${\mathcal O}(1)$. 
The resulting branching fractions for $b\to ss\bar d$ and $b\to dd\bar s$ decays, assuming positive 
interference between SM and NP contributions, are collected in Table \ref{table:resultsNP}. 
For $b\to ss\bar d$ decays the NP and SM contributions are roughly of the same size, while for $b\to dd\bar s$ 
the NP induced branching ratios are more than two orders of magnitude larger than the SM ones. 
This means that with enough statistics one could probe flavor violation without theoretical uncertainty 
to scales $\Lambda\sim 10^3$ TeV both in $3\to 2$ and $3\to 1$ transitions and not just in $2\to 1$ 
transitions as is possible now from $K-\bar K$ mixing. Of course, the statistics needed to achieve 
such an ambitious goal is well beyond the reach of present and planned flavor factories.

\section{NP leading to non-SM chiralities}
\label{non-SM-chiral}

We now turn to the description of effects induced by the local operators with non-standard chiralities
$Q_{2-5}, \tilde Q_{1-5}$. It is convenient to normalize the matrix elements of these operators
to the ones of the SM operator $Q_1$ 
\begin{eqnarray}\label{rdef}
r_j(B\to M_1M_2) \equiv \frac{\langle M_1 M_2 |Q_j|B\rangle}{\langle M_1M_2|Q_1|B\rangle} \,,
\end{eqnarray}
and similarly for $\tilde Q_{1-5}$, where the ratio is denoted as $\tilde r_j$. 
To obtain predictions for a $b\to ss\bar d$ decay branching ratio due to a particular NP chiral structure, 
one only needs to multiply the results in Table \ref{table:resultsNP} with appropriate $r_j^2$ or $\tilde r_j^2$.

Using parity one can relate $r_j$ and $\tilde r_j$, since 
$P^\dagger Q_j P = \tilde  Q_j$. For $B\to PP$ ($B\to VP$) decays one then has 
\beq
\tilde r_1 = \mp1, \qquad \text{and}\qquad \tilde r_j = \mp  r_j\,,\quad j = 2,\dots,5.
\eeq
For $B\to VV$ decays it is convenient to define  ratios $r_{\lambda,j}$, $\tilde r_{\lambda,j}$ for final states with 
definite helicites, $|V_{1,\lambda} V_{2,\lambda}\rangle$, where $\lambda = 0,\pm$. 
We then have $\tilde r_{1,\pm} =\tilde r_{1,0}= - 1$ and
\begin{eqnarray}
\tilde r_{j,\pm} = - r_{j,\mp}\,,\qquad
\tilde r_{j,0} = - r_{j,0}, \quad \quad j = 2, \dots, 5
\end{eqnarray}

We only need to compute the ratios $r_j$, $j=2, \dots,5$. The ratios $\tilde r_j$ are then 
already given by the above relations. To compute $r_j$ we use 
naive factorization \cite{Wirbel:1985ji}, which suffices for the accuracy required here. 
Strictly speaking, naive factorization is not valid at leading
order in the heavy quark expansion, but corresponds
to assuming dominance of the soft-overlap contributions in the
complete SCET factorization formula~\cite{Bauer:2004tj}, and keeping only terms of leading order in
$\alpha_s(m_b)$. In the QCDF approach, this corresponds to neglecting hard spectator 
scattering contributions \cite{Beneke:1999br,Beneke:2003zv}. 
If needed, these assumptions can be relaxed.

Naive factorization, or the vacuum insertion approximation, is also justified
in the $1/N_c$ expansion for the matrix elements of the operators $Q_{1,2,4}$, 
but not for $Q_{3,5}$. To see this, one can rewrite $Q_3$  as a sum of
color singlet and color octet terms using the color Fierz identity,
\begin{eqnarray}
Q_3 = (\bar s^\alpha_R b_L^\beta)(\bar s^\beta_R d^\alpha_L) = 
\frac{1}{N_c}  (\bar s_R b_L)(\bar s_R d_L) + 
2(\bar s_R  t^a b_L)(\bar s_R t^a d_L),
\end{eqnarray}
and analogously for $Q_5$. The matrix element of the color-singlet operator scales
as $N_c^{1/2}$, while that of the color-octet scales as $N_c^{-1/2}$. The two term in the above 
decomposition thus contribute at the same order in $1/N_c$ expansion, and both should in principle be kept.

For the experimentally interesting $B\to PV$ and $B\to VV$ decay modes all the 
ratios can be expressed in terms of $r_{2,4}$. One has $r_{3,5} = 3 r_{2,4}$,
and $r_4 = -1/[2(N_c+1)]$. 

The ratio $r_2$ is common to all the $PV$ modes which depend only on the
graphical amplitudes $t_P+c_P$ (for which the spectator quark ends up in the pseudoscalar
meson) and is given by
\begin{eqnarray}
r_2 = \frac{1}{8(N_c+1)} \frac{f_V^\perp f_T(m_V^2)}{f_V f_+(m_V^2)}
\frac{2(m_B^2-m_P^2-m_V^2)}{m_V(m_B+m_P)}\,.
\end{eqnarray}
Using $f_{K^*} = 218$ MeV, $f_{K^*}^\perp=175$ MeV and the form factors from Ref.~\cite{Ball:1998tj} 
we find $r_2(K^+K^{*0}) = 0.28$ and $r_2(\pi^+\bar K^{*0}) = 0.27$.

For the $VV$ modes we quote only the ratios corresponding to longitudinally polarized vector
mesons, which dominate the total rate. We find $r_4^\parallel = -1/[2(N_c+1)]$ and
\begin{eqnarray}
r_2^{\parallel} = - \frac{1}{8(N_c+1)} \frac{3f_{V1}^\perp}{f_{V1} m_{V1}}
\frac{4m_B^2 \vec p^2 T_1(m_{V1}^2) - (m_B^2-m_{V2}^2)(m_B^2-m_{V1}^2-m_{V2}^2) T_2(m_{V1}^2)}{(m_B+m_{V2})(m_B^2-m_{V1}^2-m_{V2}^2)A_1(m_{V1}^2)-\frac{4m_B^2\vec p^2}{m_B+m_{V2}}A_2(m_{V1}^2)}.\,
\end{eqnarray}
Here $V_1$ denotes the neutral $K^*$ meson ($K^{*0}$ for $b\to ss\bar d$ transitions, 
and $\bar K^{*0}$ for the $b\to dd\bar s$ transitions), if $V_1, V_2$ are different vector mesons.
Numerically we find
\begin{eqnarray}
&& r_2^\parallel(B^+\to K^{*+} K^{*0}, K^{*0} K^{*0}) = 0.002\,,\\
&& r_2^\parallel(B^+\to \bar K^{*0} \rho^+) = 10^{-5}\,,\qquad
r_2^\parallel(B_s\to \bar K^{*0} \bar K^{*0}) = 0.002\,,\nonumber
\end{eqnarray}
where we used the $B\to V$ form factors from Ref.~\cite{Ball:1998kk}.

\section{Conclusions}
The exclusive rare $B$ decays $b\to ss\bar d$ and 
$b\to dd\bar s$ analyzed in this paper appear in the SM only at second order in the weak interactions and have thus
very small branching fractions,
but in NP models they can be greatly enhanced. We construct the complete effective Hamiltonian
contributing to these modes in the SM, and point out the presence of nonlocal
contributions, not included in previous work, which can contribute about 30\% of the
local term. 

We show that the hadronic matrix elements of the local operators contributing
to these exclusive decays in the SM can be determined using SU(3) flavor symmetry in terms
of measured $\Delta S = 0$ decay amplitudes. Detailed numerical predictions are
given for all $B\to PP, VP, VV$ modes of experimental interest, both in the SM and for
several examples of  NP models: NP with conserved global charge, (N)MFV models and general flavor violating models. 

A general NP mechanism was identified which can enhance the branching fractions of these modes, while obeying existing constraints on NP in $\Delta S=2$ mixing processes. This mechanism represents a generalization of the sneutrino exchange in R parity violating SUSY. Any observation of such a decay mode gives a constraint on the ratio of flavor couplings to the NP, and can 
exclude regions in the parameter space of the NP theory for branching fractions observable
at LHC-b and super-B factories.

\acknowledgements{We thank B. Golob, S.~Fajfer, S. Jaeger, and A. Weiler for comments and discussions.
D.P. thanks the CERN Theory Division for hospitality during the
completion of this work.}

\appendix

\section{The structure of the effective Hamiltonian}
\label{app:1}

Consider for definiteness the 
$b\to ss\bar d$ transitions. At scales $M_W > \mu > m_b$, these transitions are described
by an effective Hamiltonian with propagating $u,c$ quarks. Writing explicitly the quarks
propagating in the box diagram, and not assuming the unitarity of the
CKM matrix, the effective Hamiltonian is given by
\begin{eqnarray}
{\cal H}_{ss\bar d} = \lambda_t^d \lambda_t^b {\cal H}(t,t) + \sum_{q_1,q_2 = u,c} \lambda_{q_1}^d \lambda_{q_2}^b
{\cal H}(q_1, q_2),
\end{eqnarray}
with $\lambda_q^{q'}=V_{qq'}V_{qs}^*$ (so that for instance $\lambda_t^d=V_{td}V_{ts}^*$. The top term in the Hamiltonian is a local operator
\begin{eqnarray}
{\cal H}(t,t) = \frac{G_F^2 m_W^2}{2} C(m_t^2/m_W^2, \mu/m_W) [(\bar s d)_{V-A} (\bar s b)_{V-A}],
\end{eqnarray}
where $C(m_t^2/m_W^2, \mu/m_W)$ is a Wilson coefficient.
The box diagram with internal quarks $q_1, q_2 = u,c$, on the other hand, is matched onto an effective Hamiltonian containing both
local and nonlocal terms \cite{Witten:1976kx} 
\begin{eqnarray}\label{H12}
{\cal H}(q_1,q_2) &=& \frac{G_F^2}{2}\left\{ A\Big(\frac{\mu}{m_W}\Big) m_W^2 [ (\bar s d)_{V-A} (\bar s b)_{V-A}] +
B\Big(\frac{\mu}{m_W}\Big) (m_1^2 + m_2^2)  [ (\bar s d)_{V-A} (\bar s b)_{V-A}] \right. \nonumber\\
& & \hspace{-2cm} \left. +
D\Big(\frac{\mu}{m_W}\Big) m_b^2   [ (\bar s d)_{V-A} (\bar s b)_{V-A}] \right\} +
\int d^dx \,T\big\{{\cal H}_d(q_2,q_1)(x), {\cal H}_b(q_1,q_2)(0)\big\}\,.
\end{eqnarray}
The effective Hamiltonian ${\cal H}_b(q_1,q_2)$ mediates $b\to s q_1 \bar q_2$ transitions, and is given by
\beq\label{H_bq1q2}
 {\cal H}_b(q_1,q_2)=\frac{G_F}{\sqrt2}\Big(\sum_{i=1,2}C_i Q_{i,b}^{q_1q_2} + \delta_{q_1 q_2}\sum_{j=3}^6 C_jQ_j^b\Big).
 \eeq
${\cal H}_d(q_1,q_2)$ mediates $d\to s \bar q_1 q_2$ and is given by a similar expression, with the 
replacement $b\to d$.
Note that there is no top-charm contribution at leading order in the $m_i^2/m_W^2$ expansion. 
The top-charm box is matched in the effective theory onto six-quark operators of the form
$(\bar c b)(\bar s c)(\bar sd)$, which are power suppressed by $1/m_W^2$ relative to the 4-quark operators shown.
Such terms appear only after using the unitarity of the CKM matrix.

The dependence on the light quark masses $m_{1,2}$ in the effective theory expression 
Eq.~(\ref{H12})
can be obtained in the mass insertion approximation. The $W^\pm$ coupling $W^+_\mu (\bar u_L \gamma_\mu d_L)$ conserves
chirality, which implies that only $m_1^2, m_2^2$
terms are allowed, but not $m_1 m_2$, which would require one mass insertion on each
propagating line. The $m_b^2$ term arises from two mass insertions on the incoming $b$ quark
line. This term is not present in $K^0-\bar K^0$ mixing. 
On the other hand, in a theory with chiral-odd quark couplings, such as e.g.
the charged Higgs couplings $H^+(\bar u_L d_R)$ in the 2HDM, another term can appear
in Eq.~(\ref{H12}), proportional to $m_1 m_2$. Chirality prevents also the appearance of
terms of the form $m_b m_1, m_b m_2$. 

Under renormalization,  the local operator with Wilson coefficient $A(\mu/m_W)$ renormalizes
multiplicatively, while the nonlocal operators mix into the local operators with coefficients
$B(\mu/m_W), D(\mu/m_W)$.

Making use of the unitarity of the CKM matrix, it is possible to eliminate $\lambda_u^b, \lambda_u^d$
as $\lambda_u^i = - \lambda_c^i-\lambda_t^i$, $i=b,d$. This reproduces the effective Hamiltonian 
quoted in text Eq.~(\ref{DeltaS=2}). The terms proportional to $A,C$ and $D$ are combined into $C_{tt}$, while the $B$ term in Eq.~(\ref{H_bq1q2}) reproduces the $C_{tc}$ and $C_{ct}$ coefficients.
The total contribution of the local terms proportional to the Wilson coefficient $B(\mu/m_W)$ is equal to
\begin{eqnarray}
& &\lambda_u^d \lambda_u^b \cdot 0 + 
\lambda_c^d \lambda_u^b \cdot m_c^2 + 
\lambda_u^d \lambda_c^b \cdot m_c^2 + 
\lambda_c^d \lambda_c^b \cdot 2m_c^2 \\
& & = -\lambda_c^d (\lambda_c^b + \lambda_t^b) \cdot m_c^2 
-\lambda_c^b (\lambda_c^d + \lambda_t^d) \cdot m_c^2 + 
\lambda_c^d \lambda_c^b \cdot 2m_c^2 = - (\lambda_c^d \lambda_t^b + \lambda_c^d \lambda_t^b) m_c^2\,.
\nonumber
\end{eqnarray}
This proves the two properties of the local effective Hamiltonian 
${\cal H}^{\Delta S = 2}$ stated in the text: i) the equality $C_{tc} = C_{ct}$, and
ii) the absence of a $\lambda_c^d \lambda_c^b$ local term. The latter property does not
hold in the presence of chiral-odd quark couplings, as for example in the 2HDM as 
discussed above.

\section{$\Delta S=2$ Wilson coefficients}\label{Wilson_coeffs}

In this appendix we show the translation of results obtained for $\bar K^0-K^0$ mixing to the 
case of $b\to ss\bar d$ decays (the results for $b\to dd\bar s$ decays are equivalent). The results for  
$\bar K^0-K^0$ mixing were derived in \cite{Gilman:1982ap} in the leading-log approximation, and in 
\cite{Herrlich:1996vf} in the next-to-leading log approximation.

We start with the Wilson coefficient $C_{tt}$, which is obtained by matching the $u,c,t$ loops at the
weak scale onto the local operator $(\bar s b)_{V-A} (\bar s d)_{V-A}$. Below this scale, QCD radiative
corrections introduce a correction $\eta_2(\mu)$, so that at NLO
\begin{eqnarray}
{C}_{tt}(\mu) = \eta_2(\mu) S_0(x_t) + \frac{1}{8\pi^2} \frac{m_b^2}{m_W^2} D(\mu/m_W).
\end{eqnarray}
The box function $S_0(x_t)$ with $x_t=m_t^2/M_W^2$ is the same as obtained in the one-loop
matching at the $m_W$ scale for $\bar K^0-K^0$ mixing (external $b$ quark leg can be considered 
as massless for the purpose of this calculation). It is given by \cite{Buchalla:1995vs} 
\begin{eqnarray}
S_0(x_t) = \frac{4x_t - 11x_t^2 + x_t^3}{4(1-x_t)^2} - \frac{3x_t^3 \log x_t}{2(1-x_t)^3}=2.26.
\end{eqnarray}
with the numerical value given for $\bar m_t(\bar m_t)=160.9$ GeV. 
The QCD correction $\eta_2(\mu)$ is obtained by solving the renormalization group equation
\begin{eqnarray}
\mu \frac{d\eta_2(\mu)}{d\mu} = \gamma_{+} \eta_2(\mu)
\end{eqnarray}
At one-loop order, the anomalous dimension is $\gamma_{+} = \alpha_s/\pi$, which gives using $\alpha_S(m_Z)=0.118$ (from which $\Lambda_{\overline{\rm MS}}^{n_f=5}=226$ MeV)
\begin{eqnarray}
\eta_2(\mu_b) = \left( \frac{\alpha_s(M_W)}{\alpha_s(m_b)}\right)^{6/23}
&=& 0.85\,, \qquad m_b=4.2 \mbox{ GeV},
\end{eqnarray}
so that 
\beq
C_{tt}(\mu_b)=1.92.
\eeq

The coefficient $D(\mu)$ parameterizes the $b$ quark mass effects, and
is introduced by mixing from the nonlocal operators into the local
operator $m_b^2 (\bar s b)_{V-A} (\bar s d)_{V-A}$. This mixing has not been computed yet.
We will neglect this contribution since it is suppressed by the small ratio $m_b^2/m_W^2\sim 0.2\%$.

The $\lambda_t^b \lambda_t^d$ nonlocal contributions due to insertions of two four-quark operators 
are power suppressed and can be neglected as discussed in appendix \ref{app:1}. 
This is no longer true for top-charm contributions, where both local and nonlocal contributions are 
power suppressed by $m_i^2/m_W^2$, and mix under renormalization.  

We use the derivation of \cite{Herrlich:1996vf}, which we adapt to the $b\to ss\bar d$ process at hand. The local
part of the  $\bar K^0-K^0$ mixing weak Hamiltonian for $\mu$ above the charm quark mass (i.e. before charm quark 
is integrated out) is given by \cite{Herrlich:1996vf}
\beq\label{KKmix}
H_{\rm eff}^{\bar K-K}=\frac{G_F^2}{2}\lambda_c^d\lambda_t^d\tilde C_7 \tilde Q_7,\qquad
\tilde Q_{7}=\frac{m_c^2}{g^2}[(\bar s d)_{V-A}(\bar s d)_{V-A}].
\eeq
The corresponding local part of the $b\to ss\bar d$  effective Hamiltonian on the other hand is
\beq\label{bssdbar}
\frac{G_F^2 m_W^2}{16 \pi^2}\big(\lambda_c^d\lambda_t^b C_{ct}+\lambda_t^d\lambda_c^b C_{tc}\big)\big[(\bar s d)_{V-A} (\bar s b)_{V-A}\big],
 \eeq
The RG evolution  calculation for $b\to ss\bar d$ process is the same as for $\bar K^0-K^0$ mixing, except that the total contribution is split into two because of two different  CKM element structures in \eqref{bssdbar}. As shown in Appendix \ref{app:1}, these structures have identical coefficients
in the SM $C_{ct}=C_{tc}$.

The same equality can be seen also in the anomalous dimension matrices for the running of these
coefficients. Consider the nonlocal contribution to $b\to ss\bar d$ with insertions of the tree operators $T\{Q_{1,2}Q_{1,2}\}$, which is given by
\beq
\label{tree-T}
\sum_{i,j=1,2} C_iC_j \Big\{ \lambda_c^d\lambda_t^b\big(Q_{i,d}^{uu}Q_{j,b}^{uu}-Q_{i,d}^{cu}Q_{j,b}^{uc}\big) +\lambda_t^d\lambda_c^b\big(Q_{i,d}^{uu}Q_{j,b}^{uu}-Q_{i,d}^{uc}Q_{j,b}^{cu}\big)\Big\}.
\eeq
When computing the mixing into the local operator $\tilde Q_7$, 
the terms in the first and the second brackets give the same contributions, since the quark 
masses are not relevant for the calculation of the anomalous dimensions (it does not matter 
whether $c$ quark or $u$ quark runs in the lower leg of the loop in Fig \ref{fig:matching}). 
This shows that the RG running for $C_{ct}, C_{tc}$ is the same. Furthermore, this running
is the same as that of $\tilde C_7$ in $K^0-\bar K^0$ mixing. This can be seen by comparing
(\ref{tree-T}) with the nonlocal operator contributing to $\bar K^0-K^0$ mixing 
\beq
\begin{split}
\sum_{i,j=1,2} C_iC_j & \lambda_c^d\lambda_t^d\big(2Q_{i,d}^{uu}Q_{j,d}^{uu}-Q_{i,d}^{cu}Q_{j,d}^{uc}-Q_{i,d}^{cu}Q_{j,d}^{uc}\big),
\end{split}
\eeq 
The two operators are identical,  provided that one sets $b\to d$ in \eqref{tree-T}. The same correspondence
between $K^0-\bar K^0$ mixing and $b\to ss\bar d$ applies also for the nonlocal contributions involving penguin operators.

In conclusion, comparing the Eqs.~(\ref{KKmix}) and (\ref{bssdbar}) we find that for $\mu>m_c$, we have
\beq
C_{ct}(\mu) = C_{tc}(\mu) = \tilde C_7(\mu) x_c \pi/\alpha_s\,,
\eeq
 where $\tilde C_7(\mu)$ is obtained from 
RG evolution in the same way as for $\bar K^0 -K^0$ mixing. A very compact form of RG equations was presented in \cite{Herrlich:1996vf}
\beq
\mu \frac{d}{d\mu}\vec D=\hat\gamma^T\cdot \vec D,
\eeq
with $\hat \gamma$ the $8\times 8$ anomalous dimension, given in Eqs. (6.23)-(6.26) and (12.50)-(12.56) of  \cite{Buchalla:1995vs}
and 
\beq
\vec D^T=(\vec C^T, C_{7+}/C_+,C_{7-}/C_-).
\eeq
Here $\vec C$ is a vector of $C_i$, $i=1,\dots 6$, $C_\pm=C_1\pm C_2$, \footnote{Here we caution about the definition of $C_{1,2}$,
 which differs from the one in  \cite{Buchalla:1995vs}. We use the definition, where $C_1(\mu_W)\sim 1$, $C_2(\mu_W)\sim 0$.}
and $\tilde C_7$ was split to $\tilde C_7=C_{7+}+C_{7-}$, where the distribution between $C_{7+}$ and $C_{7-}$ is arbitrary. At LO
we have for the matching at weak scale $\vec D^T(\mu_W)=(1,0,0,0,0,0,0,0)$, so that the nonzero value of $\tilde C_7(\mu)$ comes
entirely from the running, from mixing with $C_1$. At $\mu_b$ the solution of RG running at LO is
\beq
\vec D(\mu)=V\Big(\Big[\frac{\alpha_S(m_W)}{\alpha_S(\mu)}\Big]^{\vec \gamma^{(0)}/2\beta_0}\Big)_D V^{-1},
\eeq
with $ \hat \gamma=\frac{\alpha_S}{4\pi}\gamma^{(0)}$ and $V$ a matrix that diagonalizes the LO anomalous dimension matrix, $\gamma_D^{(0)}=V^{-1}\gamma^{(0)T}V$. This gives
\beq
\tilde C_7(m_b)=0.268, \quad m_b=4.2~{\rm GeV},
\eeq
 and finally
 \beq
\tilde C_{tc}(m_b)=3.75 x_c=9.35 \cdot 10^{-4},
\eeq
where in the last equality we used $m_c=1.27$ GeV. 

\section{Bounds on the flavor-changing couplings}\label{app:bounds}
We have showed in the introduction that $b\to ss\bar d$ branching ratios can be large, if NP effects are due 
to exchange of particle(s) with conserved charge. The resulting effective weak Hamiltonian, Eq. \eqref{NPX}, depends
on four couplings, $g_{s\to d}, g_{d\to s}, g_{b\to s}, g_{s\to b}$  and an overall mass scale $M_X$, that in this appendix we set to $M_X=10$ TeV (this then fixes the overall normalization of $g_i$). In order to have large $b\to ss\bar d$ branching ratios and simultaneously avoid bounds from $K-\bar K$ mixing and $B_s-\bar B_s$ mixing a hierarchy between couplings is required. Another way of looking at this is that, if a large $b\to ss\bar d$ decay branching ratio (we will quantify what "large" means below) is found by Belle~II and/or LHCb this would imply that a region of parameter space with $g_{s\to b} \sim g_{s\to d}$ and/or $g_{b\to s}\sim g_{d\to s}$ would be excluded. We show this below.

The experimental constraints from $K-\bar K$ mixing and $B_s - \bar B_s$ 
mixing give the following upper bounds (fixing $M_X=10$ TeV and using bounds from Eq. \eqref{bounds})
\begin{eqnarray}\label{rewrite:bounds}
\varepsilon_{sd}\equiv |g_{d\to s}g_{s\to d}^*| \leq \frac{M_X^2}{\Lambda_{sd}^2} =10^{-4}\,,\qquad
\varepsilon_{bs}\equiv|g_{b\to s}g_{s\to b}^*| \leq \frac{M_X^2}{\Lambda_{bs}^2} =0.11 \,.
\end{eqnarray}
We also define the following two ratios of coupling constants
\begin{eqnarray}\label{R12def}
R =  \frac{g_{s\to b }}{g_{s\to d}}\,,\qquad
\bar R=  \frac{g_{b\to s}}{g_{d\to s}}\,.
\end{eqnarray}
We now show that a measured {\em lower} bound on the $b\to ss\bar d$ branching fraction 
excludes values of $R, \bar R$ that are close to 1. For definiteness, we assume that the NP field $X$ couples to the quarks with the
Dirac structure $\Gamma = P_R$, as in RPV SUSY. Similar bounds can be derived
for any other Dirac structure $\Gamma$. 

The amplitude for the $\bar B\to f$ transition mediated by the operator
 $(\bar s b)(\bar s d)$, Eq.~(\ref{NPX}), is
 \beq\label{ampg}
A(\bar B\to f) = \frac{1}{M_X^2} \langle f |
 g_{d\to s}g_{s\to b}^* Q_4 +g_{b\to s}g_{s\to d}^* \tilde Q_4|\bar B\rangle =
\frac{r_4}{M_X^2} \langle f |Q_1|\bar B\rangle 
(g_{d\to s}g_{s\to b}^* \mp g_{b\to s}g_{s\to d}^*),
\eeq
where the upper (lower) sign is for a $PP (PV)$ final state.
The combination of couplings $g_i$ can be written in terms of the ratios
$R, \bar R$ defined in (\ref{R12def})
\beq
g_{b\to s}g_{s\to d}^* \mp g_{d\to s}g_{s\to b}^* = 
g_{ b\to s}g_{s\to b}^* \frac{1}{R^*} \mp g_{d\to s}g_{s\to d}^* R^* 
= 
(g_{d\to s}g_{s\to d}^*) \bar R \mp (g_{b\to s}g_{s\to b}^*) \frac{1}{\bar R}\,.
\eeq
The products of coefficients on the r.h.s are now exactly the ones bounded from the meson mixing, Eq. \eqref{rewrite:bounds}.
The absolute value of the l.h.s on the other hand is assumed to be bounded from below from the measurement of $b \to ss\bar d$ 
branching ratio, cf. Eq. \eqref{ampg}. We then have
\beq
B^2<|g_{b\to s}g_{s\to d}^* \mp g_{d\to s}g_{s\to b}^*|^2 
\leq \varepsilon_{sd}^2 |R|^2 + \varepsilon_{bs}^2 \frac{1}{|R|^2} + 
2\varepsilon_{sd} \varepsilon_{bs}\,.
\eeq
If $B \geq 2\sqrt{\varepsilon_{sd} \varepsilon_{bs}}$, then the above inequality rules out a range of values for $|R|$,
\beq\label{boundR1}
\frac{1}{2\varepsilon_{sd}^2} [B - 2\varepsilon_{sd}\varepsilon_{bs} - \sqrt{B^2-4\varepsilon_{sd}\varepsilon_{bs}}]
\leq |R|^2 \leq 
\frac{1}{2\varepsilon_{sd}^2} [B - 2\varepsilon_{sd}\varepsilon_{bs} + \sqrt{B^2-4\varepsilon_{sd}\varepsilon_{bs}}].
\eeq
The same bound with $\varepsilon_{sd}\leftrightarrow \varepsilon_{bs}$ holds also for $|\bar R|$. 
The requirement $B \geq 2\sqrt{\varepsilon_{sd} \varepsilon_{bs}}$ corresponds to the requirement 
that ${\cal B}(B\to f)>4 {\cal B}(B\to f)_{\rm NMFV}$, with the 
NMFV predictions for branching ratios given in Table \ref{table:resultsNP}.


\begin{thebibliography}{39}

\bibitem{Huitu:1998vn}
  K.~Huitu, D.~X.~Zhang, C.~D.~Lu and P.~Singer,
  Phys.\ Rev.\ Lett.\  {\bf 81}, 4313 (1998)
  [arXiv:hep-ph/9809566].

\bibitem{Grossman:1999av}
  Y.~Grossman, M.~Neubert and A.~L.~Kagan,
  JHEP {\bf 9910}, 029 (1999)
  [arXiv:hep-ph/9909297].


\bibitem{Fajfer:2000ny}
  S.~Fajfer and P.~Singer,
  Phys.\ Rev.\  D {\bf 62}, 117702 (2000)
  [arXiv:hep-ph/0007132].

\bibitem{Fajfer:2006av}
  S.~Fajfer, J.~F.~Kamenik and N.~Kosnik,
  Phys.\ Rev.\  D {\bf 74}, 034027 (2006)
  [arXiv:hep-ph/0605260].
\bibitem{Fajfer:2001ht}
  S.~Fajfer and P.~Singer,
  Phys.\ Rev.\  D {\bf 65}, 017301 (2002)
  [arXiv:hep-ph/0110233].
  
\bibitem{Cai:2004mi}
  H.~Y.~Cai and D.~X.~Zhang,
  Commun.\ Theor.\ Phys.\  {\bf 44}, 887 (2005)
  [arXiv:hep-ph/0410144].
\bibitem{Fajfer:2004fx}
  S.~Fajfer, J.~F.~Kamenik and P.~Singer,
  Phys.\ Rev.\  D {\bf 70}, 074022 (2004)
  [arXiv:hep-ph/0407223].
  
\bibitem{Wu:2003kp}
  X.~H.~Wu and D.~X.~Zhang,
  Phys.\ Lett.\  B {\bf 587}, 95 (2004)
  [arXiv:hep-ph/0312177].


\bibitem{Bona:2007vi}
  M.~Bona {\it et al.}  [UTfit Collaboration],
  JHEP {\bf 0803}, 049 (2008)
  [arXiv:0707.0636 [hep-ph]].

\bibitem{Kagan:2009bn}
  A.~L.~Kagan, G.~Perez, T.~Volansky and J.~Zupan,
  arXiv:0903.1794 [hep-ph].

\bibitem{Buchalla:1995vs}
  G.~Buchalla, A.~J.~Buras and M.~E.~Lautenbacher,
  Rev.\ Mod.\ Phys.\  {\bf 68}, 1125 (1996)
  [arXiv:hep-ph/9512380].

\bibitem{Witten:1976kx}
  E.~Witten,
  Nucl.\ Phys.\  B {\bf 122}, 109 (1977).

\bibitem{Gilman:1982ap}
  F.~J.~Gilman and M.~B.~Wise,
  Phys.\ Rev.\  D {\bf 27}, 1128 (1983).

\bibitem{Herrlich:1996vf}
  S.~Herrlich and U.~Nierste,
  Nucl.\ Phys.\  B {\bf 476}, 27 (1996)
  [arXiv:hep-ph/9604330].
  
\bibitem{Fajfer:2000ax}
  S.~Fajfer and P.~Singer,
  Phys.\ Lett.\  B {\bf 478}, 185 (2000)
  [arXiv:hep-ph/0001132].
  
\bibitem{Charles:2004jd}
  J.~Charles {\it et al.}  [CKMfitter Group],
  Eur.\ Phys.\ J.\  C {\bf 41}, 1 (2005)
  [arXiv:hep-ph/0406184]; we are using the spring 2009 update available at http://ckmfitter.in2p3.fr.

\bibitem{Gronau:1998fn}
  M.~Gronau, D.~Pirjol and T.~M.~Yan,
  Phys.\ Rev.\  D {\bf 60}, 034021 (1999)
  [Erratum-ibid.\  D {\bf 69}, 119901 (2004)]
  [arXiv:hep-ph/9810482].

\bibitem{Gronau:1994rj}
  M.~Gronau, O.~F.~Hernandez, D.~London and J.~L.~Rosner,
  Phys.\ Rev.\  D {\bf 50}, 4529 (1994)
  [arXiv:hep-ph/9404283].

\bibitem{Grinstein:1996us}
  B.~Grinstein and R.~F.~Lebed,
  Phys.\ Rev.\  D {\bf 53}, 6344 (1996)
  [arXiv:hep-ph/9602218].

\bibitem{Dighe:1997wj}
  A.~S.~Dighe, M.~Gronau and J.~L.~Rosner,
  Phys.\ Rev.\  D {\bf 57}, 1783 (1998)
  [arXiv:hep-ph/9709223].

\bibitem{Gronau:2000az}
  M.~Gronau,
  Phys.\ Rev.\  D {\bf 62}, 014031 (2000).
  
\bibitem{HFAG}
  E.~Barberio {\it et al.}  [Heavy Flavor Averaging Group (HFAG)
                  Collaboration],
  arXiv:0808.1297 [hep-ex], summer 2009 update.
  
\bibitem{Garmash:2003er}
  A.~Garmash {\it et al.}  [Belle Collaboration],
  Phys.\ Rev.\  D {\bf 69}, 012001 (2004)
  [arXiv:hep-ex/0307082].

\bibitem{Aubert:2008rr}
  B.~Aubert {\it et al.}  [BABAR Collaboration],
  Phys.\ Rev.\  D {\bf 78}, 091102 (2008)
  [arXiv:0808.0900 [hep-ex]].


\bibitem{Browder:2008em}
  T.~E.~Browder, T.~Gershon, D.~Pirjol, A.~Soni and J.~Zupan,
  arXiv:0802.3201 [hep-ph].
  
\bibitem{Kusaka:2007mj}
  A.~Kusaka {\it et al.}  [Belle Collaboration],
  Phys.\ Rev.\  D {\bf 77}, 072001 (2008)
  [arXiv:0710.4974 [hep-ex]].

\bibitem{Aubert:2006fg}
  B.~Aubert {\it et al.}  [BABAR Collaboration],
  arXiv:hep-ex/0608002.




\bibitem{D'Ambrosio:2002ex}
  G.~D'Ambrosio, G.~F.~Giudice, G.~Isidori and A.~Strumia,
  Nucl.\ Phys.\  B {\bf 645}, 155 (2002)
  [arXiv:hep-ph/0207036].

\bibitem{Agashe:2005hk}
  K.~Agashe, M.~Papucci, G.~Perez and D.~Pirjol,
  arXiv:hep-ph/0509117.

\bibitem{Wirbel:1985ji}
  M.~Wirbel, B.~Stech and M.~Bauer,
  Z.\ Phys.\  C {\bf 29}, 637 (1985).
  
\bibitem{Bauer:2004tj}
  C.~W.~Bauer, D.~Pirjol, I.~Z.~Rothstein and I.~W.~Stewart,
  Phys.\ Rev.\  D {\bf 70}, 054015 (2004)
  [arXiv:hep-ph/0401188];
  C.~W.~Bauer, I.~Z.~Rothstein and I.~W.~Stewart,
  Phys.\ Rev.\  D {\bf 74}, 034010 (2006)
  [arXiv:hep-ph/0510241];
  A.~R.~Williamson and J.~Zupan,
  Phys.\ Rev.\  D {\bf 74}, 014003 (2006)
  [arXiv:hep-ph/0601214];
\bibitem{Chay:2003zp}
  J.~g.~Chay and C.~Kim,
  Phys.\ Rev.\  D {\bf 68}, 071502 (2003)
  [arXiv:hep-ph/0301055].
\bibitem{Beneke:1999br}
  M.~Beneke, G.~Buchalla, M.~Neubert and C.~T.~Sachrajda,
  Phys.\ Rev.\ Lett.\  {\bf 83}, 1914 (1999)
  [arXiv:hep-ph/9905312].

\bibitem{Beneke:2003zv}
  M.~Beneke and M.~Neubert,
  Nucl.\ Phys.\  B {\bf 675}, 333 (2003)
  [arXiv:hep-ph/0308039].
\bibitem{Ball:1998tj}
  P.~Ball,
  JHEP {\bf 9809}, 005 (1998)
  [arXiv:hep-ph/9802394].

\bibitem{Ball:1998kk}
  P.~Ball and V.~M.~Braun,
  Phys.\ Rev.\  D {\bf 58}, 094016 (1998)
  [arXiv:hep-ph/9805422].



\end{thebibliography}
\end{document}